\documentclass[11pt,a4paper]{article}

\pdfoutput=1
\usepackage{jheppub}
\usepackage{amsfonts}
\usepackage{amsmath}
\usepackage{amssymb}
\usepackage{graphicx,color}
\usepackage{float}
\usepackage{hyperref}
\usepackage{subfigure}
\graphicspath{{Figures/},{Logos/}}
\usepackage{slashed}
\usepackage{tikz}
\usetikzlibrary{arrows,shapes}
\usetikzlibrary{trees}
\usetikzlibrary{matrix,arrows} 				
\usetikzlibrary{positioning}				
\usetikzlibrary{calc,through}				
\usetikzlibrary{decorations.pathreplacing}  
\usepackage{pgffor}							

\usetikzlibrary{decorations.pathmorphing}	
\usetikzlibrary{decorations.markings}
\tikzset{
	vector/.style={decorate, decoration={snake}, draw},
	provector/.style={decorate, decoration={snake,amplitude=2.5pt}, draw},
	antivector/.style={decorate, decoration={snake,amplitude=-2.5pt}, draw},
	fermion/.style={draw=black, postaction={decorate},
		decoration={markings,mark=at position .55 with {\arrow[draw=black]{>}}}},
	fermionbar/.style={draw=black, postaction={decorate},
		decoration={markings,mark=at position .55 with {\arrow[draw=black]{<}}}},
	fermionnoarrow/.style={draw=black},
	gluon/.style={decorate, draw=black,
		decoration={coil,amplitude=4pt, segment length=5pt}},
	scalar/.style={dashed,draw=black, postaction={decorate},
		decoration={markings,mark=at position .55 with {\arrow[draw=black]{>}}}},
	scalarbar/.style={dashed,draw=black, postaction={decorate},
		decoration={markings,mark=at position .55 with {\arrow[draw=black]{<}}}},
	scalarnoarrow/.style={dashed,draw=black},
	electron/.style={draw=black, postaction={decorate},
		decoration={markings,mark=at position .55 with {\arrow[draw=black]{>}}}},
	bigvector/.style={decorate, decoration={snake,amplitude=4pt}, draw},
}

\tikzstyle{block} = [draw, rectangle, 
minimum height=3em, minimum width=6em]

\title{$\nu$-Inflaton Dark Matter}

\author[a]{Ant\'onio Torres Manso}   
\author[b]{Jo\~{a}o G.~Rosa} 

\affiliation[a]{Departamento F\'isica e Astronomia da Faculdade Ci\^encias da Universidade do Porto, Porto}

\affiliation[b]{Departamento de F\'isica da Universidade de Aveiro and CIDMA,  Campus de Santiago, 3810-183 Aveiro, Portugal}

\emailAdd{up201303871@fc.up.pt}
\emailAdd{joao.rosa@ua.pt}

\abstract{
We present a unified model where the same scalar field can drive inflation and account for the present dark matter abundance. This scenario is based on the incomplete decay of the inflaton field into right-handed neutrino pairs, which is accomplished by imposing a discrete interchange symmetry on the inflaton and on two of the right-handed neutrinos. We show that this can lead to a successful reheating of the Universe after inflation, while leaving a stable inflaton remnant at late times. This remnant may be in the form of WIMP-like inflaton particles or of an oscillating inflaton condensate, depending on whether or not the latter evaporates and reaches thermal equilibrium with the cosmic plasma. We further show that this scenario is compatible with generating light neutrino masses and mixings through the seesaw mechanism, predicting at least one massless neutrino, and also the observed baryon asymmetry via thermal leptogenesis.}


\begin{document}
\maketitle

\section{Introduction}

The inflationary paradigm, developed in the 1980's \cite{Guth:1980zm,Linde:1981mu,Albrecht:1982wi}, with an introduction of an early phase of accelerated expansion prior to the radiation domination epoch, is expected to solve the major problems of Standard Cosmology, in particular the flatness, horizon and the large scale structure puzzles. The early period of accelerated expansion would flatten the universe and stretch the initial small inhomogeneities to superhorizon scales, that later evolved into the present large scale structures. In addition, the initial causal contact between regions that are presently too far to have interacted with each other would be ensured, reconnecting the theory to the almost perfect isotropy of the Cosmic Microwave Background (CMB). However, this solution is yet to be reconciled with the Standard Model (SM) of particle physics.

Similarly, a common problem to both particle physics interactions and the cosmological behavior is the energy budget of the Universe. Only around 25$\%$ of the Universe is made of matter \cite{Tyson:1998vp,Dahle:2007wf}, and only about 15$\%$ of this is accounted for by the SM particles \cite{Paczynski:1985jf}, leaving the remaining part, referred to as dark matter, to be described by new physics (see e.g.~\cite{Taoso:2007qk}).

From experimental results with neutrino oscillations, we know neutrinos to be massive, but the mechanism for mass generation is unknown \cite{Gariazzo:2018pei}. The SM only includes left-handed neutrinos, with three generations of them, as opposed to the other fermions, which have also a right-handed component. Going beyond the SM, though without violating its structural symmetries, and allowing for the existence of right-handed neutrinos, also in three generations, could explain the mass appearance through what is known as the seesaw mechanism \cite{Mohapatra:2006gs,Gouvea:2016shl}. Another important open question in cosmology is the observed asymmetry between particles and anti-particles \cite{Sakharov:1967dj,Cline:2006ts,Davidson:2008bu}. Assuming an inflationary evolution, diluting any initial asymmetry in the Universe, the observed asymmetry should have been generated dynamically after inflation. Such mechanisms must follow the conditions derived by Sakharov \cite{Sakharov:1967dj}, but within the SM no mechanisms are able to generate the required amount of asymmetry \cite{Cline:2006ts}, again requiring extensions to new physics. Nevertheless, the introduction of right-handed neutrinos presents a viable way to solve this puzzle through leptogenesis.

Expecting both the inflaton and dark matter to be weakly interacting neutral fields, and outside the reach of the present framework of the SM, it is interesting to speculate whether the same field can describe both at different epochs. Scalar fields, depending on the kinematical regimes, may behave as fluids with different equations of state. For example, in a regime of a slowly varying field, $\dot{\phi}^2 /2 \ll V(\phi)$, it acts as an effective cosmological constant, whereas while oscillating about the minimum of the potential it may behave as either nonrelativistic matter or as relativistic matter for quadratic and quartic potentials, respectively. Since these regimes are commonly present in an inflationary description, a unification model seems plausible. The general problem in an unified picture lies in the connection to the standard cosmological evolution before the light elements generation in Big Bag Nucleosynthesis (BBN). Although an effective reheating must be ensured, it may be possible to have an incomplete inflaton decay, leaving room for a stable remnant that mimics dark matter behavior, as discussed in several examples in the literature \cite{Panotopoulos:2007ri,Cardenas:2007xh,Liddle:2008bm,Okada:2010jd,Bose:2009kc,DeSantiago:2011qb,Lerner:2009xg,Akhmedov:1998qx,Khoze:2013uia,Borah:2018rca}. It would also be interesting if such a unification scenario could be embedded within the simplest required extension of the SM including right-handed neutrinos, allowing for light neutrino masses and leptogenesis.

In this work, we develop an inflationary model, regulated by a non-minimal coupling between gravity and the inflaton scalar field, where the latter may decay only into two fermions, which we identify with two of the right-handed neutrinos\footnote{For other examples of scenarios where neutrinos are coupled to the dark sector, see e.g.~\cite{Bernardini:2007hm,Bernardini:2008pn}.}. By imposing a discrete symmetry in the inflaton-right-handed neutrino sector, along the lines originally proposed in \cite{Bastero-Gil:2015lga}, we ensure that no other inflaton decays are possible, such that when the decay into right-handed neutrinos becomes kinematically forbidden one is left with a stable inflaton remnant that can account for the present dark matter abundance. We thus refer to this scenario as the $\nu$-Inflaton Dark Matter ($\nu$IDM) model.

The introduction of right-handed neutrinos, as decay products of the inflaton, and their interaction with the SM particles through Yukawa couplings then allows for the generation of the observed light neutrino masses and mixings through the seesaw mechanism. The remnant scalar field may survive as an oscillating condensate or evaporate into thermalized inflaton particles in equilibrium with the cosmic plasma, depending on the scattering rate with the radiation bath particles. In the latter case inflaton particles will follow a decoupling-freeze out dynamics typical of Weakly Interacting Massive Particles (WIMP), and in both scenarios the inflaton remnant will behave as pressureless, non-relativistic or cold dark matter (CDM). Finally, we analyze how the observed baryon asymmetry may be generated through thermal leptogenesis, namely via the decays of the third right-handed neutrino, which does not result directly from inflaton decay but may nevertheless be thermally produced during/after reheating. 

This work is organized as follows. In the next section we introduce our model by  detailing the imposed symmetry, its properties and consequences, the model's Lagrangian and its most important features. In section \ref{sec3} we explore the consequences of the imposed discrete symmetry on the light neutrino masses and mixings. The dynamics of our model is detailed in section \ref{sec4}, where we describe inflation, the reheating period, and the evolution of the scalar inflaton particles as WIMPs or as an oscillating condensate, depending on the parametric regime. Finally, in section \ref{sec5} we embed a thermal leptogenesis scenario in our model scenario. In section \ref{sec6} we summarize and discuss our main results.


\section{$\nu$IDM model description}

We consider a real scalar field, the inflaton field $\phi$, with a potential energy $V(\phi)$ and a non-minimal coupling to gravity of the form $\xi\phi^2 R/2$. The inflaton interacts with two right-handed neutrinos, $N_{1}$ and $N_{2}$, through standard Yukawa interactions, with an imposed discrete interchange symmetry $C_{2}\subset\mathbb{Z}_{2}\times S_{2}$, under which the inflaton field and the fermion fields transform as:
\begin{equation}
\phi  \leftrightarrow-\phi~, \qquad 
N_{1} \leftrightarrow N_{2}~.
\end{equation}
We also consider an additional right-handed neutrino, $N_3$, that does not interact with the inflaton, so as to match the number of fermion generations in the SM. All three right-handed neutrinos are SM singlet Weyl fermions, allowing for Majorana mass terms, and interact with the Higgs and lepton doublets through standard Yukawa interactions, which allows for a seesaw mechanism to generate light neutrino masses as we detail below. 

The complete Lagrangian for our model is the following:
\begin{equation}
\mathrm{\mathsf{\mathcal{L}}}=\mathsf{\mathcal{L}}_{\phi}+\mathsf{\mathcal{L}}_{N\,Kin}+\mathsf{\mathcal{L}}_{N\,Mass}+\mathsf{\mathcal{L}}_{SM\leftrightarrow N}~,
\end{equation}
in addition to the SM Lagrangian. The inflation sector describes the scalar field kinetic and potential terms as well as its non-minimal coupling to gravity, along with the Einstein-Hilbert gravitational term:
\begin{equation}
\mathsf{\mathcal{L}}_{Inf}=\frac{M^{2}+\xi\phi^{2}}{2}R-\frac{1}{2}\partial_{\mu}\phi\partial^{\mu}\phi-V(\phi)~,\label{Eq:IDM Inflation L}
\end{equation}
where $M$ is a mass parameter, $R$ is the Ricci scalar and $\xi$ is a dimensionless coupling. Notice that if $\xi=0$ we have the standard minimal coupling to gravity and $M=M_{P}$, the reduced Planck mass. The parameter $\xi$ turns the effective Planck mass into a dynamical quantity. We consider the most general renormalizable scalar potential compatible with the $\mathbb{Z}_2$ reflection symmetry $\phi \leftrightarrow -\phi$:
\begin{equation}
V(\phi)=\frac{1}{2}m_{\phi}^{2}\phi^{2}+\lambda\phi^{4}~.
\label{Eq:IDM Inflation Pot}
\end{equation}
Although both a quadratic and a quartic potential potential are in tension with the state-of-the-art Planck limits \cite{Ade:2015xua} on the scalar spectral index, $n_s$, and tensor-to-scalar ratio, $r$, it is well known that the inclusion of a non-minimal coupling to gravity makes the Einstein-frame scalar potential effectively plateau-like and akin to the Starobinsky inflationary model based on the addition of $R^2$ terms to the Einstein-Hilbert action \cite{Starobinsky:1980te}, in which case both $n_s$ and $r$ can be in very good agreement with Planck data  \cite{Ade:2015xua}\footnote{Note that other inflationary scenarios may also accommodate a quartic scalar potential in agreement with Planck data, as in the case of warm inflation (see e.g.~\cite{Bartrum:2013fia, Bastero-Gil:2016qru}).}.

As we will see, the quartic term is expected to be dominant during inflation since $\lambda\gtrsim 10^{-10}$ and $m_{\phi}^{2}\lesssim1\ \mathrm{TeV}$ in the interesting parametric range, with the inflaton taking superplanckian values during inflation. However, with just a quartic term the field would be massless at the origin and could not behave as cold dark matter at late times, which justifies the inclusion of both quadratic and quartic terms.

The second term on the right-hand side of the Lagrangian includes the kinetic terms of the three right-handed neutrinos and their Majorana mass terms are given by the third term, which also includes the Yukawa interactions with the inflaton field:
\begin{equation}
\mathsf{\mathcal{L}}_{N\,Mass}=-\frac{1}{2}(M_{1}+h\phi)N_{1}N_{1}^{c}-\frac{1}{2}(M_{2}-h\phi)N_{2}N_{2}^{c}-\frac{1}{2}M_{3}N_{3}N_{3}^{c}~.
\end{equation}
Note that the interchange symmetry imposes $M_1=M_2$ and that the inflaton couples equally to $N_1$ and $N_2$, although with an opposite sign, which is crucial for ensuring its stability at late times. Notice, in addition, that in the light neutrino mass generation through the seesaw mechanism, which occurs after the spontaneous breaking of the electroweak gauge symmetry, the inflaton field  will lie close to the minimum at the origin, leaving only the bare masses in the Majorana mass matrix. 

Finally, there are the Yukawa interactions
\begin{equation}
\mathsf{\mathcal{L}}_{SM\leftrightarrow N}=y_{\mathtt{i\mathtt{\ell}}}N_{i}^{c}H\,L+h.c.\,,\label{Eq:SM rhneutrino interaction term}	
\end{equation} 
involving the three right-handed neutrinos, the Higgs doublet and the SM left-handed lepton doublets. In the Higgs mechanism, the Higgs fields acquires a vacuum expectation value, $\left\langle H\right\rangle =\frac{v}{\sqrt{2}}\left(0, 1\right)^t$, thus generating a Dirac mass term for the neutrinos. Restrictions coming from neutrino oscillation experiments will then constrain our model and limit its allowed parametric range.

The imposed discrete symmetry plays a major role in the reheating period and in ensuring that a stable inflaton remnant is left at late times. In particular, it forbids all linear couplings of the inflaton to any fields apart from $N_1$ and $N_2$, such that at the minimum of the potential at $\phi=0$ (assuming no spontaneous breaking of the symmetry), the only possible inflaton decays are into $N_1$ and $N_2$ pairs. Thus, for $M_1> m_\phi/2$ the inflaton is stable at the minimum.  However, while the inflaton oscillates about the origin the $\mathbb{Z}_2$ symmetry is broken and the effective Majorana masses $M_\pm = |M_1\pm h\phi|$ can take values below $m_\phi/2$, such that the decay into $N_1$ and $N_2$ pairs is permitted for a range of field values. This then allows for a successful reheating after inflation while making the inflaton decay incomplete, as we will analyze in detail in section 4.

\begin{figure}[H]

\centering 

\begin{tikzpicture}[line width=1 pt, scale=1,baseline=0]

\draw[fermionbar] (0:0)--(-40:1);
\draw[fermion] (0:0)--(40:1);
\draw[scalar] (180:1)--(0:0);

\draw[fermionbar,blue] (-40:1)--(40:1);
\draw[scalar,blue] (-40:1)--(2,-0.64);
\draw[scalar,blue] (40:1)--(2,0.64);

\end{tikzpicture}
\hspace{1cm}
\begin{tikzpicture}[line width=1 pt, scale=1,baseline=0]

\draw[fermionbar] (0:0)--(-40:1);
\draw[fermion] (0:0)--(40:1);
\draw[scalar] (180:1)--(0:0);

\draw[scalarbar,blue] (-40:1)--(40:1);
\draw[fermion,blue] (-40:1)--(2,-0.64);
\draw[fermion,blue] (40:1)--(2,0.64);

\end{tikzpicture}
\hspace{1cm}
\begin{tikzpicture}[line width=1 pt, scale=1,baseline=-0.3]

\draw[fermion] (0:0)--(-40:1);
\draw[fermion] (0:0)--(40:1);
\draw[scalar] (180:1)--(0:0);

\draw[scalar,blue] (-40:1)--(-5:1.7);
\draw[scalar,blue] (40:1)--(5:1.7);
\draw[fermion,blue] (-40:1)--(-30:2);
\draw[fermion,blue] (40:1)--(30:2);...

\end{tikzpicture}
	
	\caption{ Examples of inflaton decay channels forbidden by the discrete interchange symmetry. For clarity, we represent the light Higgs and neutrino fields in blue and the inflaton and right-handed neutrinos in black. } \label{fig:FDecaysIDM}
\end{figure}
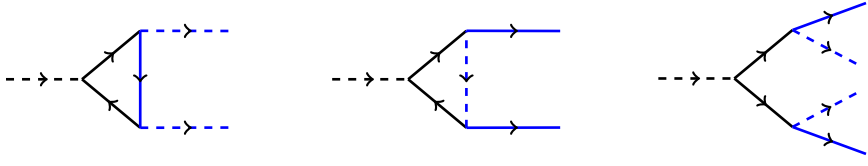

Note the importance of the imposed symmetry for blocking inflaton decay modes involving off-shell right-handed neutrinos, thus allowing for a stable inflaton remnant below the kinematic threshold. In particular, as illustrated in figure \ref{fig:FDecaysIDM}, the contributions of virtual $N_1$ and $N_2$ modes to the inflaton decay into other light particles cancel each other, due to their opposite sign coupling to the inflaton field.

\section{Seesaw mechanism in the $\nu$IDM scenario}
\label{sec3}

Before we discuss the dynamics of the model, let us consider the impact of imposing the discrete interchange symmetry on the spectrum of light neutrino masses at late times, when the inflaton is lying close to the minimum, $\phi \ll M_1/h$. The relevant Lagrangian for the seesaw mechanism is thus: 
\begin{equation}
\mathcal{L}_{mass}=-\frac{1}{2}M_{1}N_{1}N_{1}^{c}-\frac{1}{2}M_{2}N_{2}N_{2}^{c}-\frac{1}{2}M_{3}N_{3}N_{3}^{c}
+y_{\mathtt{i\mathtt{\ell}}}N_{i}^{c}H\,\nu_{\mathtt{\mathtt{\ell}}}+h.c.
\end{equation}
with $\mathtt{\ell}=e,\mu,\tau$ and $i=1,2,3$, and the interchange symmetry imposes $M_1=M_2$ and $y_{\mathtt{1\mathtt{\ell}}}=y_{\mathtt{2\mathtt{\ell}}}$.

Let us look at the resulting mass matrices. Since we have the freedom to consider a diagonal Majorana mass matrix, then
\begin{equation}
M_{R}=\left(\begin{array}{ccc}
M_{1} & 0 & 0\\
0 & M_{1} & 0\\
0 & 0 & M_{3}
\end{array}\right).
\end{equation}
As a result of the $N_{1}\leftrightarrow N_{2}$ symmetry,  the Dirac mass matrix after electroweak symmetry breaking can be written as:
\begin{equation}
m_{D}={v\over \sqrt{2}}\begin{pmatrix}y_{1e} & y_{1\mu} & y_{1\tau}\\
y_{2e} & y_{2\mu} & y_{2\tau}\\
y_{3e} & y_{3\mu} & y_{3\tau}
\end{pmatrix}={v\over \sqrt{2}}\begin{pmatrix}a & b & c\\
a & b & c\\
d & e & f
\end{pmatrix}~,
\end{equation}
where we have defined:
\begin{eqnarray}
a=y_{1e}=y_{2e}~, & \quad b=y_{1\mu}=y_{2\mu}~, &\quad c=y_{1\tau}=y_{2\tau}~,\nonumber \\
d=y_{3e}~,&\quad e=y_{3\mu}~, &\quad f=y_{3\tau}~.
\end{eqnarray}
Upon diagonalization, the light neutrino mass matrix can then be written as \cite{Mohapatra:2006gs,Gouvea:2016shl}:
\begin{equation}
\mathcal{M}_{\nu}=-m_{D}^{T}M_{R}^{-1}m_{D}\,=-{v^2\over 2}\begin{pmatrix}\frac{d^{2}}{M_{3}}+\frac{2a^{2}}{M_{1}} & \frac{de}{M_{3}}+\frac{2ab}{M_{1}} & \frac{df}{M_{3}}+\frac{2ac}{M_{1}}\\
\frac{de}{M_{3}}+\frac{2ab}{M_{1}} & \frac{e^{2}}{M_{3}}+\frac{2b^{2}}{M_{1}} & \frac{ef}{M_{3}}+\frac{2bc}{M_{1}}\\
\frac{df}{M_{3}}+\frac{2ac}{M_{1}} & \frac{ef}{M_{3}}+\frac{2bc}{M_{1}} & \frac{f^{2}}{M_{3}}+\frac{2c^{2}}{M_{1}}
\end{pmatrix}~.\label{Eq:seesaw IDM}
\end{equation}
It is then easy to check that, as a result of the interchange symmetry, $\det\mathbf{\,\mathcal{M}}_{\nu}=0$ and at least one of the light neutrinos must be massless.

On the other hand, since the flavour basis charged lepton mass matrix
is diagonal, the mass matrix of the light neutrinos, in the same basis,
is diagonalized by $U_{PMNS}$ through
\begin{equation}
M_{\nu}^{d}=U_{PMNS}^{T}\mathcal{M}_{\nu}U_{PMNS}~.
\end{equation}
Thus, we obtain 
\begin{equation}
\mathbf{\mathcal{M}}_{\nu}=U_{PMNS}^{*}M_{\nu}^{d}U_{PMNS}^{\dagger}~,
\end{equation}
where 
\begin{equation}
M_{\nu}^{d}=\begin{pmatrix}m_{1} & 0 & 0\\
0 & m_{2}\mathit{e}^{-2i\phi_{1}} & 0\\
0 & 0 & m_{3}\mathit{e}^{-2i\phi_{2}}
\end{pmatrix}~.
\end{equation}
As a first approximation, we will not consider the Majorana phases and, using neutrino oscillation data on $\Delta m_{12}^{2}=\left|m_{1}^{2}-m_{2}^{2}\right|\sim10^{-5}\ \mathrm{eV^{2}}\ll\Delta m_{23}^2$ \cite{Gariazzo:2018pei}, we may consider $m_{1}\simeq m_{2}$ to leading order. As discussed in the literature (see e.g.~\cite{Mohapatra:2006gs}), the tribimaximal mixing matrix yields a very good approximation to the experimental data on neutrino mixing angles:
\begin{equation}
U_{PMNS}\simeq U_{TBM}=\frac{1}{\sqrt{6}}\begin{pmatrix}2 & \sqrt{2} & 0\\
-1 & \sqrt{2} & \sqrt{3}\\
1 & -\sqrt{2} & \sqrt{3}
\end{pmatrix}~.
\end{equation}
Thus, we have in the tribimaximal approximation:
\begin{equation}
\mathcal{M}_{\nu}=\begin{pmatrix}m_{2} & 0 & 0\\
0 & \frac{m_{2}+m_{3}}{2} & -\frac{m_{2}-m_{3}}{2}\\
0 & -\frac{m_{2}-m_{3}}{2} & \frac{m_{2}+m_{3}}{2}
\end{pmatrix}~.\label{Eq:mass with UTBM IDM}
\end{equation}
Matching the two light neutrino mass matrices we may then obtain conditions on the parameters of the model.
Using equations \eqref{Eq:seesaw IDM} and \eqref{Eq:mass with UTBM IDM} we obtain two distinct solutions.
The first corresponds to a spectrum with inverted hierarchy:
\begin{eqnarray}\label{Eq:solution 1 n mass IDM}
m_1\simeq m_2=\frac{(y_{1e}^{2}+2y_{1\mu}^{2})v^2}{M_{1}}~,\qquad m_3=0\nonumber\\
y_{1e}=\pm y_{3\mu}\gamma^{\frac{1}{2}}~, \qquad y_{1\mu}=\mp\frac{y_{3e}}{2}\gamma^{\frac{1}{2}}~, \qquad y_{i\tau}=-y_{i\mu}
\end{eqnarray}
where $i=1,2,3$ and $\gamma=M_1/M_3$, which clearly exhibits a $\mu\leftrightarrow \tau$ symmetry, such that interactions with $\nu_{\mu}$ have an opposite sign to interactions with $\nu_{\tau}$.

Using now the experimental value for $m_2^2\simeq \Delta m_{23}^{2}=(2.44\pm0.06)\times\mathrm{10^{-3}\ eV^{2}}$, we find:
\begin{equation}
M_{1}=1.21\times10^{15}\,y_{eff}^{2}\mathrm{\ GeV},
\label{Eq:M1}
\end{equation}
where we have defined $y_{eff}^2 =\sum_{\ell}y_{1\ell}^{2}$. In the second solution, we find a normal neutrino hierarchy: 
\begin{eqnarray}
&m_1\simeq m_2= 0~, \qquad m_{3}=v^{2}\left(\frac{2y_{1\mu}^{2}}{M_{1}}+\frac{y_{3\mu}^{2}}{M_{3}}\right)\nonumber\\
&y_{ie}=0;\quad y_{i\tau}=y_{i\mu}~,
\label{Eq:solution 2 n mass IDM}
\end{eqnarray}
which again exhibits a $\mu\leftrightarrow\tau$ symmetry, that is in fact a feature of the tribimaximal mixing. Due to $U_{PMNS}=U_{TBM}$ and since $m_{2}\simeq m_{1}= 0$ there are no Yukawa interactions with $\nu_{e}$, i.e.~they should be very suppressed once deviations from tribimaximal mixing are taken into account.

In this second scenario we cannot write the non-vanishing neutrino mass only in terms of parameters related to $N_1$ and $N_2$, which are directly involved in the reheating process. Nevertheless, provided that the contributions of $N_1$ and $N_3$ to the eigenvalue $m_3$ are comparable, a relation of the form (\ref{Eq:M1}) will also hold up to $\mathcal{O}(1)$ numbers after imposing the experimental value for $m_3^2 \simeq \Delta m_{23}^{2}$ as in the first case.

In our analysis of the cosmological dynamics, we will then use Eq.~(\ref{Eq:M1}) to relate the Majorana mass $M_1$ and the effective Yukawa coupling $y_{eff}$ so as to obtain an experimentally consistent neutrino mass spectrum at low energies, up to small deviations from tribimaximal mixing that will not significantly affect our discussion.


\section{Cosmological dynamics of the $\nu$IDM scenario}
\label{sec4} 

\subsection{Inflation}
\label{sub4.1}

As mentioned earlier, the simple quadratic and quartic potentials cannot give an observationally consistent description of the inflationary period, which may be overcome by considering a non-minimal coupling to gravity as we review in this subsection.

Writing the action with the already presented inflation Lagrangian, Eq. \eqref{Eq:IDM Inflation L}
\begin{equation}
S_{J}=\int d^{4}x\sqrt{-g}\left[\frac{M^{2}+\xi\phi^{2}}{2}R-\frac{1}{2}\partial_{\mu}\phi\partial^{\mu}\phi-V(\phi)\right]~,
\label{Eq:non-minimalaction}
\end{equation}
we see that the non-minimal coupling turns the effective Planck mass into a dynamical
quantity. The present Planck mass is equal to the mass parameter $M$, since $\left\langle \phi\right\rangle=0 $ at late times.

\begin{figure}[H]
	\centering
	\includegraphics[totalheight=6cm]{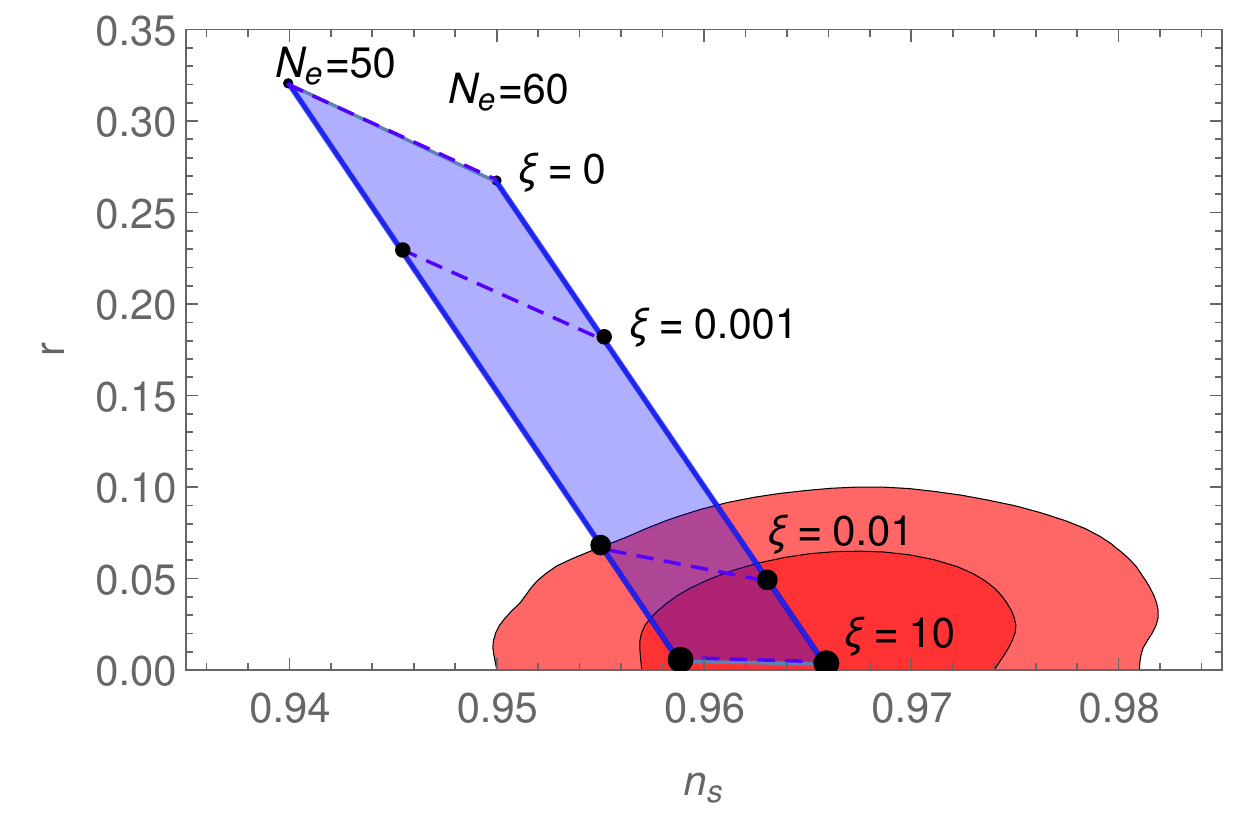}
	\caption{ Observational predictions for an inflaton non-minimally coupled to gravity with a quartic potential for 50-60 e-folds of inflation. The red contours correspond to the 68\% and 95\% C.L.~results from Planck.} \label{fig:NMC}
\end{figure}

The inflationary dynamics is better analyzed in the Einstein frame, which can be obtained by performing a conformal transformation of the form\footnote{See e.g. \cite{Postma:2014vaa} ~for a discussion on the equivalence of using the Jordan and Einstein frames for studying the inflationary dynamics.}:
\begin{equation}
\hat{g}_{\mu\nu}=\Omega^{2}g_{\mu\nu},\qquad\Omega^{2}=1+\frac{\xi\phi^{2}}{M_{P}^{2}}~.
\end{equation}
The Einstein frame Ricci scalar is then given by 
\begin{equation}
R=\Omega^{2}\left[\hat{R}-6\left(\nabla^{2}\ln\Omega+\left(\nabla\ln\Omega\right)^{2}\right)\right]~.
\end{equation}
This transformation leads to a non-minimal kinetic term for the inflaton field. To obtain the canonical form it is convenient to perform a
field redefinition $\phi\rightarrow\chi$ such that:
\begin{equation}
\frac{d\chi}{d\phi}=\sqrt{\frac{\Omega^{2}+6\xi^{2}\phi^{2}/M_{P}^{2}}{\Omega^{4}}}~.
\end{equation}
After this redefinition, the action in the Einstein frame becomes
\begin{equation}
S_{E}=\int d^{4}x\sqrt{-\hat{g}}\left[\frac{M_{P}^{2}}{2}\hat{R}-\frac{1}{2}\partial_{\mu}\chi\partial^{\mu}\chi-U(\chi)\right]~,
\end{equation}
where $U(\chi)$ is just a conformal rescaling of the Jordan frame potential:
\begin{equation}
U(\chi)=\Omega^{-4}V(\phi)~.
\end{equation}
If $V(\phi)$ is dominated by the quartic term, we then obtain a nearly flat plateau at large field values, $\phi\gg M_P/\sqrt{\xi}$, with $U(\chi)\simeq \lambda M_P^4/\xi^2$. Having written the action in the canonical Einstein-Hilbert form allows for a standard analysis of inflation in the slow-roll regime, with the slow-roll parameters and number of e-folds of inflation given by:
\begin{equation}
\epsilon=\frac{1}{2}M_{p}^{2}\left(\frac{U'(\chi)}{U(\chi)}\right)^{2}~,\qquad\eta=M_{p}^{2}\frac{U''(\chi)}{U(\chi)}~,\qquad N_e=\frac{1}{M_{P}^{2}}\int_{\chi_{e}}^{\chi_{*}}\frac{U(\chi)}{U'(\chi)}d\chi~,\label{Eq:slowroll IDM}
\end{equation}
where $\chi_*$ and $\chi_e$ denote the field value when the relevant CMB scales exit the horizon during inflation and at the end of inflation, respectively. Neglecting the quadratic contribution at large field values, we then obtain:
\begin{flalign}
\epsilon & =\frac{8M_{P}^{4}}{\phi^{2}\left[M_{P}^2+\xi(1+6\xi)\phi^{2}\right]} \label{Eq:epsilon}~,\\
\eta& =\frac{12M_{P}^{6}+4M_{P}^{4}\xi(1+12\xi)\phi^{2}-8M_{P}^{2}\xi^{2}(1+6\xi)\phi^{4}}{\phi^2\left[M_{P}^{2}+\xi(1+6\xi)\phi^{2}\right]^{2}}~, \label{Eq:Eta}\\
N_e & =\frac{1}{8}\left[\frac{(1+6\xi)\phi^{2}}{M_{P}^{2}}-6\ln\left[M_{P}^{2}+\xi\phi^{2}\right]\right]\Biggl|_{\phi_{e}}^{\phi_{*}}\simeq\frac{(1+6\xi)\phi_{*}^{2}}{8M_{P}^{2}}~.
\end{flalign}
To generate the observed amplitude for the spectrum of scalar curvature perturbations $\Delta_{{\cal R}}^{2}=2.2\text{\ensuremath{\times}}10^{-9}$, we obtain the condition $U/\epsilon_{\chi}=(0.0269M_{P})^{4}$ \cite{Bezrukov:2007ep}. This yields a relation between the non-minimal coupling and the inflation self-interaction:
\begin{equation}
\lambda \simeq 4\times 10^{-10}\xi^2.\label{Eq:nmc and lambda}
\end{equation}
The scalar to tensor ratio, $r$, and the scalar spectral index $n_{s}$ are given by:
\begin{eqnarray}
r & =& 16\epsilon = \frac{16(1+6\xi)}{N_{e}+8N_{e}^{2}\xi}\simeq {12\over N_e^2}\nonumber\\
n_{s}&= & 1-6\epsilon+2\eta =\frac{-3+N_{e}-18\xi-40N_{e}\xi+16N_{e}^{2}\xi-192N_{e}\xi^{2}-128N_{e}^{2}\xi^{2}+64N_{e}^{3}\xi^{2}}{N_{e}+16N_{e}^{2}\xi+64N_{e}^{3}\xi^{2}}\nonumber\\
&\simeq& 1-{2\over N_e}~,
\end{eqnarray}
where in the last expressions we give the leading results for large $\xi$, which coincide with the predictions of the Starobinsky model  \cite{Starobinsky:1980te} and Higgs inflation \cite{Bezrukov:2007ep}. In Figure \ref{fig:NMC} we illustrate the predictions for this model in the $(n_s,r)$ plane as a function of $\xi$, for $N_{e}=50-60$ e-folds of inflation.
 
The Planck data thus imposes a lower bound on the non-minimal coupling and consequently, through Eq. (\ref{Eq:nmc and lambda}), on the inflaton self-coupling. In particular, the 1-$\sigma$ bound gives, for $N_{e}=60$, $\xi\gtrsim0.008$ and $\lambda\gtrsim3\times10^{-14}$.

At the end of the slow-roll regime, when $\epsilon \simeq 1$, we have $\phi_e\simeq M_P/\sqrt{\xi}$, which implies that the effects of the non-minimal coupling become sub-dominant right after inflation and the inflaton will start to oscillate about an essentially quartic potential. As we will see in more detail below, the inflaton's energy density begins to redshift as radiation and its decay eventually leads to a bath of relativistic particles that become the dominant component in the Universe. Hence, until matter-radiation equality the Universe is always dominated by a fluid redshifting as radiation, thus yielding $R\simeq 0$ so that we may safely neglect the effects of the non-minimal coupling to gravity in the field's dynamics. In addition, after the inflaton remnant takes over once again as cold dark matter, the field amplitude is already well below $M_P/\sqrt{\xi}$. We will thus neglect the effects of the non-minimal coupling to gravity in our analysis of the post-inflationary dynamics of the inflaton field.



\subsection{Reheating }
\label{sub4.2}

After the inflation period, we must have first a radiation dominated epoch, in which light nuclei are genereated in what is known as Big Bang Nucleosynthesis, and then the matter domination phase. Therefore, the reheating process assumes a crucial role in the theory. There is also the possibility of a transition due to non-perturbative effects, mostly in the so called preheating phase. Nevertheless, such processes are in general not sufficient to reheat the Universe, i.e.~to convert most of the energy in the inflaton field into radiation, and for this reason we will leave them out of this discussion (see also \cite{Bastero-Gil:2015lga}). 

Once the field exits the slow-roll regime, it will start oscillating about the origin, initially with a large amplitude $\sim M_P/\sqrt{\xi}$ that will redshift with expansion. This means that reheating will occur while the potential is dominated by the quartic term, and only when its oscillation amplitude falls below $\sim m/\sqrt{\lambda}$ will it oscillate in an approximately quadratic potential. Nevertheless, the analysis of the reheating period is very similar to that performed in the generic model of \cite{Bastero-Gil:2015lga} for a quadratic potential.

\begin{figure}[htbp]
	\centering
		\includegraphics[width=0.49\textwidth]{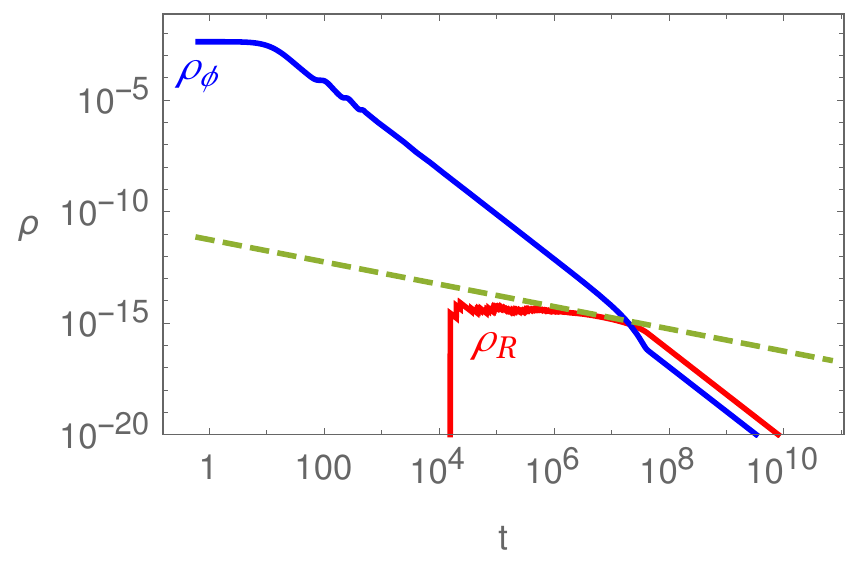}
		\includegraphics[width=0.5\textwidth]{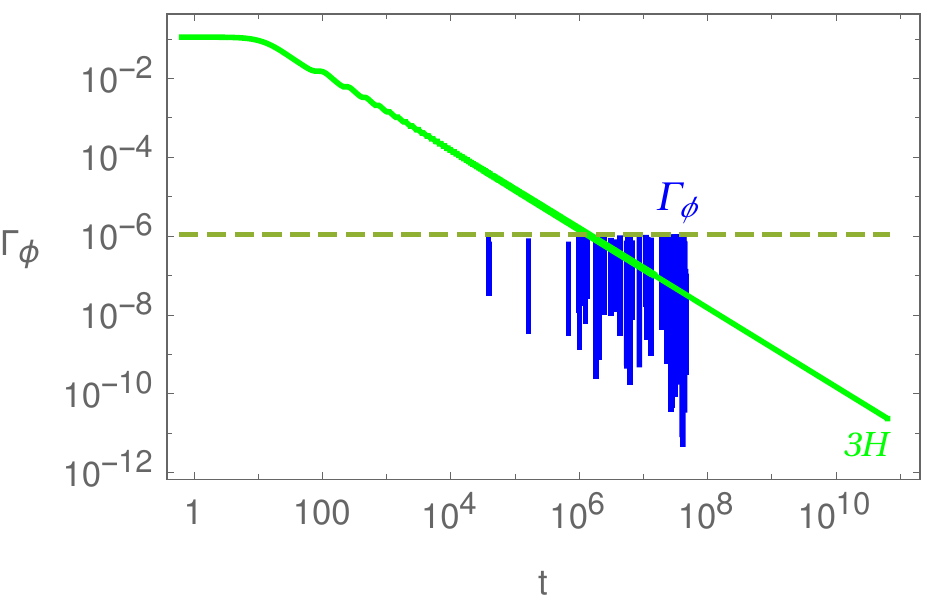}
	\caption{Numerical solution of the coupled inflaton and radiation equations of motion for $M_{1}=5\times 10^{-5}M_P$, $h=1$ and $\lambda=10^{-3}$. In the left plot we show the time evolution of the inflaton and radiation energy density, the dashed green line representing the analytical behaviour derived in Eq.~(\ref{Eq:phi^4 radiation energy density}). In the right plot we give the comparison between the Hubble parameter and the inflaton decay width, the dashed line giving the maximum value of the decay width per oscillation period. Results are given in Plank units. } 	
	\label{fig:phi4}
\end{figure}

The inflaton begins oscillating once its effective mass, $m_{eff}=\sqrt{12\lambda\phi^{2}}$, becomes larger than the Hubble parameter, its equation of motion being given by:
\begin{equation}
\ddot{\phi}+3H\dot{\phi}+\Gamma_{\phi}\dot{\phi}+4\lambda\phi^{3}=0~,
\end{equation}
where $\Gamma_\phi$ denotes the inflaton decay width into the right-handed neutrinos $N_1$ and $N_2$. Let us assume that inflaton decays result in a bath of nearly thermalized relativistic particles, an hypothesis that we discuss in more detail below. The radiation energy density will then follow:
\begin{equation}
\dot{\rho}_{r}+4H\rho_{r}=\Gamma_{\phi}\dot{\phi}^{2}~,\label{Eq:radiation energy density evolution inflation}
\end{equation} 
along with the Friedmann equation
\begin{equation}
H^{2}=\frac{\frac{1}{2}\dot{\phi}^{2}+\lambda\phi^{4}+\rho_{r}}{3M_{P}^{2}}~.
\end{equation}
The inflaton decay $\Gamma_{\phi}$ is regulated by the imposed interchange symmetry. Far from the origin, the discrete symmetry is broken, leading to a mass splitting between the two right-handed neutrinos:
\begin{equation}
M_{\pm}=\left|M_{1}\pm h\phi\right|\,.
\end{equation}
Thus, for sufficiently large $\phi$ values the inflaton may decay into the two right-handed neutrinos while it oscillates about the minimum. In particular, the kinematic condition:
\begin{equation}
\left|M_{1}\pm h\phi\right|<\frac{m_{\phi}}{2}
\end{equation}
implies that inflaton decay is only possible for field oscillation amplitudes $\gtrsim M_1/h$. The associated decay width is given by:
\begin{equation}
\Gamma_{\pm}=\frac{h^{2}}{16\pi}\sqrt{12\lambda\phi^{2}}\left(1-\frac{4\left(M_{1}\pm h\phi\right){}^{2}}{\sqrt{12\lambda\phi^{2}}}\right)^{\frac{3}{2}}~.
\end{equation}
A numerical solution of the coupled inflaton and radiation equations of motion is presented in Figure \ref{fig:phi4}.

As can be observed in Figure \ref{fig:phi4}, after the first few oscillations the inflaton energy density decreases as $t^{-\frac{1}{2}}$, thus behaving as dark radiation as expected for oscillations about a quartic potential. In this region the inflaton decays into $N_1$ and $N_2$ in short bursts during each oscillation and redshifts with the Hubble expansion. 

Before it decays significantly, the inflaton's evolution is approximately described by a damped harmonic oscillator with varying frequency:
\begin{equation}
\ddot{\phi}+3H\dot{\phi}+\omega^{2}\phi=0\,,\label{Eq:phi^4 dp osc}
\end{equation}
where $\omega$ depends on the amplitude of field oscillations, $\Phi$. Numerically, we find that  $\omega\simeq\sqrt{3\lambda\Phi^{2}}$ provides a very good approximation, in good agreement with the exact solution for the Klein-Gordon equation for a homogeneous field with a quartic potential (see e.g.~\cite{Ichikawa:2008ne}). Thus, we obtain:  
\begin{equation}
\phi(t)\simeq\Phi(t)\sin(\omega t+\alpha),\qquad\Phi(t)=\left(\frac{3}{\lambda}\right)^{\frac{1}{4}}\sqrt{\frac{M_{p}}{2t}}\,,\label{Eq:phi^4 solution}
\end{equation}
where $\alpha$ is a phase depending on the initial conditions and $\omega$ varies in time as the field's amplitude decreases with expansion.

We now consider the radiation energy density equation \eqref{Eq:radiation energy density evolution inflation}. In each oscillation period, $\tau_{\phi}=2\pi/\omega$, the kinematic condition for inflaton decay $\left|M_{1}\pm h\phi\right|=\frac{m_{\phi}}{2}$ is valid in two occasions for each right-handed neutrino during a time $\Delta t<\tau_\phi$ and the average decay width over the oscillation period is $\simeq \frac{\Gamma_{\mathrm{max}}}{2}$, where $\Gamma_{\mathrm{max}}$ denotes its maximum value. The right-hand side of the radiation equation can then be written as 
\begin{equation}
\Gamma_{\phi}\dot{\phi}^{2}=2\,\frac{\Gamma_{\mathrm{max}}}{2}\,2\,\Delta t\,\frac{\omega}{2\pi}\,\dot{\phi}^{2}~.
\end{equation}
The maximum decay width is obtained for field values $\phi=M_{1}/h$ and the field velocity is $\dot{\phi}\sim\omega\Phi$. We can then compute the field values for which the kinematic decay condition fails:
\begin{equation}
\left| M_{1}\pm\phi h\right| =\frac{\sqrt{12\lambda\phi^{2}}}{2}~, 
\end{equation}  
and use \eqref{Eq:phi^4 solution} to obtain the time $\Delta t$ for which decays are possible during each oscillation, yielding:
\begin{equation}
\Delta t\simeq\frac{M_{1}}{\omega h^{2}\Phi}\sqrt{12\lambda}~.
\end{equation}
Thus, we have 
\begin{equation}
\Gamma_{\phi}\dot{\phi}^{2}\simeq\frac{9}{4\pi^{2}h}\lambda^{2}M^{2}\Phi^{3}\simeq\frac{3^{11/4}}{2^{7/2}\pi^{2}h}\lambda^{5/4}M_{1}^{2}M_{p}^{3/2}t^{-3/2}~.
\end{equation}
The Hubble damping term in \eqref{Eq:radiation energy density evolution inflation} can be written as 
\begin{equation}
4H\,\rho_{r}=\frac{2}{t}\rho_{r}~.
\end{equation}
We may now solve \eqref{Eq:radiation energy density evolution inflation} to obtain 
\begin{equation}
\rho_{r}=\frac{3^{7/4}}{2^{5/2}\pi^{2}h}\lambda^{5/4}M_{1}^{2}M_{p}^{3/2}t^{-1/2}=\frac{3\sqrt{3}\lambda^{3/2}M_{1}^{2}M_{p}}{4\pi^{2}h}\Phi~,\label{Eq:phi^4 radiation energy density}
\end{equation}
which provides a good approximation to the numerical solution as can be seen in Figure \ref{fig:phi4}, where this analytical solution is represented by the dashed green line. 

In this calculation we assumed that the right-handed neutrinos $N_{1}$ and $N_{2}$ redshift as radiation, but this is no trivial assumption. When produced, the right-handed neutrinos are approximately massless, but their mass quickly increases with the oscillations of $\phi$, $M_{\pm}=M_{1}\pm h\phi$, becoming non-relativistic particles. On the one hand, if $N_1$ and $N_2$ decay efficiently while non-relativistic, the thermal bath will be essentially made of their decay products, i.e.~relativistic Higgs bosons and light neutrinos.  If, on the other hand, their decay rate cannot yet compete with expansion, they will redshift as matter rather than radiation.

 Although a detailed analysis of the dynamical evolution of all particle species would give a more rigorous picture of the inflaton decay, we note that the above analysis can be performed for a generic equation of state parameter, $w$, for the fluid made up of the right-handed neutrinos and their decay products. This would only change the result in Eq.~(\ref{Eq:phi^4 radiation energy density}) by a factor $6/(4+6w)$, which is $\mathcal{O}(1)$ for either radiation or non-relativistic matter, such that it does not significantly affect our analysis. 

Radiation will then become the dominant component in the Universe, $\rho_r>\rho_\phi$, when the amplitude of inflaton oscillations falls below:
\begin{equation}
\Phi^3<\frac{3\sqrt{3}\lambda^{\frac{1}{2}}M_{1}^{2}M_{p}}{4\pi^{2}h}~.\label{Eq:phi^4 reheating condition 1}
\end{equation}
Since this must occur before the decay into right-handed neutrinos becomes kinematically forbidden, i.e. for $\Phi>\frac{M}{h}$, 
we obtain an upper bound on the right-handed neutrino mass:
\begin{equation}
M_{1}<\frac{3\sqrt{3}\lambda^{\frac{1}{2}}h^{2}M_{p}}{4\pi^{2}}\simeq 6.6\times 10^{12} \xi h^2\ \mathrm{GeV}~,
\label{Eq:M1 bounds IDM}
\end{equation}
where we have used Eq.~(\ref{Eq:nmc and lambda}). From the above results we may also compute the reheating temperature, i.e.~the radiation temperature at the point of inflaton-radiation equality, yielding:
\begin{equation}
T_{R}\simeq4\left(\frac{100}{g_{\star R}}\right)^{-1/4}\left(\frac{M_{1}}{\mathrm{TeV}}\right)^{2/3}\xi^{5/6}h^{-1/3}\ \mathrm{TeV}~,\label{Eq:phi^4 reheating T}
\end{equation}
where $g_{\star R}$ is the number of relativistic degrees of freedom at this stage.

Despite the above result showing that the reheating temperature can easily be above $100$ MeV, such that radiation dominates before Big Bang Nucleosynthesis (BBN) is supposed to take place in the standard cosmological model, we must ensure that quarks, leptons and gauge bosons are produced before BBN, which implies that the right-handed neutrinos must decay efficiently at sufficiently high temperatures.

Through the interaction term in Eq. \eqref{Eq:SM rhneutrino interaction term}, that after spontaneous symmetry breaking generates the Dirac neutrino mass, the right handed neutrinos may decay into the Higgs and lepton doublets, with a decay width:
\begin{equation}
\Gamma_{D}=\Gamma\left(N_{1,2}\rightarrow H\,L\right)+\Gamma\left(N_{1,2}\rightarrow\bar{H}\,\bar{L}\right)=y_{eff}^{2}\frac{M_{1}}{8\pi}~.
\end{equation}
It is important to note that the physical right-handed neutrino, i.e.~the heavy neutrino mass eigenstate, as mentioned in the introduction, is a Majorana fermion, therefore being its own antiparticle, $N_{1}=\bar{N_{1}}$. This will be an important detail when we study CP violation in leptogenesis. 

For a significant part of the right-handed neutrino's energy density to be transferred into the SM degrees of freedom we must ensure that
\begin{equation}
\frac{\Gamma_{D}}{H}=\frac{\sqrt{90}}{8\pi^{2}}g_{\star}^{-1/2}y_{eff}^{2}\frac{M_{1}M_{P}}{T^{2}}>1\,.
\end{equation}
Using the relation between the effective Yukawa coupling and the light neutrino masses obtained in \eqref{Eq:M1}, we obtain:
\begin{equation}
\frac{\Gamma_{D}}{H}=\frac{\sqrt{90}}{8\pi^{2}}g_{\star}^{-1/2}k\frac{M_{P}m_{\nu}}{v^{2}}\left(\frac{M_{1}}{T}\right)^{2}\sim10^{2}g_{\star}^{-1/2}\left(\frac{M_{1}}{T}\right)^{2}~.
\end{equation}
If the reheating temperature exceeds the right-handed neutrino mass threshold, $T_R> M_1$, we will have an initial thermal bath made up of relativistic $N_1$ and $N_2$ neutrinos, and the above result shows that they will decay into SM particles before they become non-relativistic. If $T_R<M_1$ the right-handed neutrinos decay before inflaton-radiation equality, and at reheating the thermal bath already included the SM fields. Hence, we must require $M_1, T_R\gtrsim 100$ MeV to ensure a successful cosmological synthesis of light elements.


\subsection{WIMPlaton dark matter scenario}
\label{sub4.3}

We have seen that the inflaton scalar field becomes stable at late times, since its decays are blocked kinematically and off-shell decay modes are forbidden by the interchange symmetry. Since the scalar field exhibits a  non-relativistic behavior while oscillating around the minimum, the inflaton becomes a stable dark matter candidate. However, we must take into account that the produced right-handed neutrinos, as well as their decay products, may scatter off the low-momentum inflaton particles in the oscillating condensate, promoting them to higher momentum states. If sufficiently fast compared to the Hubble rate, such scattering processes may eventually lead to a thermalization of inflaton particles with the overall cosmic plasma. These will eventually decouple from the latter and their abundance will ``freeze-out" as in the standard WIMP models. As in the original generic inflaton dark matter model of \cite{Bastero-Gil:2015lga}, we refer to this as the ``WIMPlaton scenario".

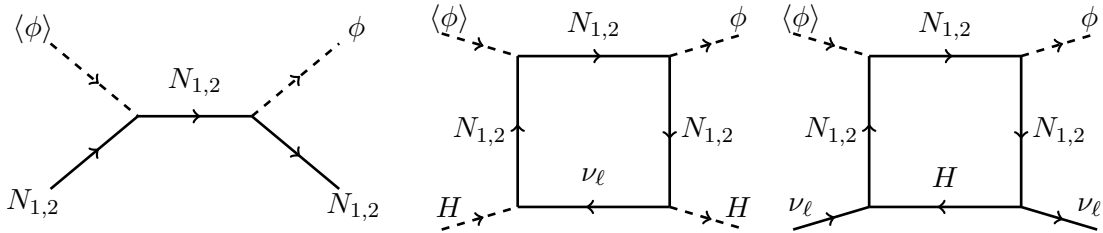
\begin{figure}[H]
	\centering
	
	\begin{tikzpicture} [line width=1 pt, scale=1.5]
	\draw[fermion] (-140:1)--(0,0);
	\draw[scalar] (140:1)--(0,0);
	\draw[fermion] (0:0)--(1,0);
	\node at (-140:1.2) {$N_{1,2}$};
	\node at (140:1.2) {$\left\langle \phi\right\rangle $};
	\node at (.5,.3) {$N_{1,2}$};	
	\begin{scope}[shift={(1,0)}]
	\draw[fermion] (0,0)--(-40:1);
	\draw[scalar] (0,0)--(40:1);
	\node at (-42:1.2) {$N_{1,2}$};
	\node at (40:1.2) {$\phi$};	
	\end{scope}
	\end{tikzpicture}\quad
	\begin{tikzpicture}[line width=1 pt, scale=1]
	\draw[scalar] (-1,2.3) -- (0,2);
	\draw[fermion] (0,2) -- (2,2);
	\draw[scalar] (2,2) -- (3,2.3);
	\node at (-0.9,2.5) {$\left\langle \phi\right\rangle $};
	\node at (2.9,2.5) {$\phi$};
	\node at (1,2.4) {$N_{1,2}$};
	
	\draw[scalar] (-1,-.3) -- (0,0);
	\draw[fermion] (2,0) -- (0,0);
	\draw[scalar] (2,0) -- (3,-.3);
	\node at (-0.9,0) {$H$};
	\node at (2.9,0) {$H$};
	\node at (1,0.4) {$\nu_{\ell}$};
	\draw[fermion] (0,0) -- (0,2);
	\draw[fermion] (2,2) -- (2,0);
	\node at (-0.5,1) {$N_{1,2}$};
	\node at (2.5,1) {$N_{1,2}$};
	\end{tikzpicture}
	\begin{tikzpicture}[line width=1 pt, scale=1]
	\draw[scalar] (-1,2.3) -- (0,2);
	\draw[fermion] (0,2) -- (2,2);
	\draw[scalar] (2,2) -- (3,2.3);
	\node at (-0.9,2.5) {$\left\langle \phi\right\rangle $};
	\node at (2.9,2.5) {$\phi$};
	\node at (1,2.4) {$N_{1,2}$};
	
	\draw[fermion] (-1,-.3) -- (0,0);
	\draw[fermion] (2,0) -- (0,0);
	\draw[fermion] (2,0) -- (3,-.3);
	\node at (-0.9,0) {$\nu_{\ell}$};
	\node at (2.9,0) {$\nu_{\ell}$};
	\node at (1,0.4) {$H$};
	\draw[fermion] (0,0) -- (0,2);
	\draw[fermion] (2,2) -- (2,0);
	\node at (-0.5,1) {$N_{1,2}$};
	\node at (2.5,1) {$N_{1,2}$};	
	
	\end{tikzpicture}
	\caption{Dominant Feynman diagrams for evaporation processes} \label{fig:Feyn1}
\end{figure}

In Figure \ref{fig:Feyn1} we illustrate the basic processes that can be responsible for the evaporation of the oscillating inflaton condensate, with $\langle \phi\rangle$ denoting a low-momentum inflaton particle in the condensate and $\phi$ a higher momentum mode after scattering. Evaporation may thus proceed through tree-level scatterings off the right-handed neutrinos or by 1-loop processes with either Higgs bosons or light neutrinos in the external states.

There is a possibility that condensate evaporation occurs while the inflaton field is still the dominant component in the energy density
of the Universe. In this scenario, the scattering processes would, nevertheless, thermalize the condensate and convert it into radiation, thus effectively yielding a successful reheating. As a consequence, reheating would happen earlier and at temperatures higher than the reheating temperature computed above.  Hence, we may take the result in Eq.~(\ref{Eq:phi^4 reheating T}) as a lower bound on the reheating temperature, and consider in more detail the case where evaporation occurs only after reheating.

Let us first consider the case where the reheating temperature is larger than $m_{\phi}$ and $M_{1}$. Proceeding as in \cite{Bastero-Gil:2015lga}, we may compute the scattering rate of low-momentum inflaton particles by right-handed neutrinos in the thermal bath, as in figure~\ref{fig:Feyn1}, assuming equilibrium distributions for the latter and also the scattered inflaton particles. The net condensate
evaporation rate, including both direct and inverse processes, is given by
\begin{multline}
\Gamma_{evap}^{(N)}=\frac{1}{n_{\phi}}\int\prod_{i=1}^{4}\frac{d^{3}\mathbf{p}_{i}}{\left(2\pi\right)^{3}2E_{i}}\left(2\pi\right)^{4}\delta^{4}\left(p_{1}+p_{2}-p_{3}-p_{4}\right)\left|{\cal M}\right|^{2}\\
\times\left[f_{1}f_{2}\left(1+f_{3}\right)\left(1-f_{4}\right)-f_{3}f_{4}\left(1+f_{1}\right)\left(1-f_{2}\right)\right]\,,
\end{multline}
where ${\cal M}$ is the scattering amplitude for the processes and $f_{i}$ are the phase space distributions for all the species involved in the initial and final states. We may take $f_1\gg 1$ for the low-momentum inflaton particles in the oscillating condensate, thus yielding to leading order for $T\gg m_{\phi},M_{\pm}$:
\begin{equation}
\Gamma_{evap}^{(N)}\simeq\frac{h^{4}}{12\pi^{3}}\left(1+\log\left(\frac{T}{m_{\phi}}\right)\right)T.
\end{equation}
In the radiation era, the evaporation rate is thus expected to catch the Hubble expansion rate, thus leading to an efficient evaporation. However, if the right-handed neutrinos become non-relativistic, the interaction rate becomes Boltzmann suppressed. Thus, evaporation must happen for $T\gtrsim M_{1}$, imposing the condition:
\begin{equation}
\frac{\Gamma_{evap}^{(N)}}{H}\Biggl|_{T=M_{1}}\simeq10^{13}h^{4}g_{\star}^{-1/2}\left(\frac{1\ \mathrm{TeV}}{M_{1}}\right)\gtrsim1~.\label{Eq:evap condition N 1}
\end{equation}
This evaporation rate is similar to the rate of other four body interactions, as inflaton and fermion annihilation. We therefore expect that, for a sufficiently large Yukawa coupling $h$, the condensate degrees of freedom are thermalized, therefore destroying the boson condensate. This conversion will increase the radiation energy density and its temperature, leading to a maximum value of the order
\begin{equation}
T_{R}^{\mathrm{max}}=\left(\frac{90}{\pi^{2}}\right)^{1/4}g_{\star}^{-1/4}\sqrt{M_{P}M_{1}}=8.5\times10^{10}g_{\star}^{-1/4}\left(\frac{M_{1}}{1\ \mathrm{TeV}}\right)^{1/2}\mathrm{GeV}~,
\end{equation}
which corresponds to the limiting case where evaporation occurs right after the beginning of the inflaton oscillations as discussed
previously. 

After annihilation and elastic scattering processes become inefficient, the thermalized inflaton particles will decouple from the thermal bath and their abundance will freeze out, as for a standard WIMP. We expect that, at this stage, both the right-handed neutrinos and the scalar particles are non-relativistic, yielding an annihilation (t-channel) cross section, given by
\begin{equation}
\sigma_{\phi\phi}\simeq\frac{h^{4}}{8\pi m_{\phi}^{2}}~.
\end{equation}
Following the standard computation of the thermal relic abundance for a decoupled species we obtain
\begin{equation}
m_{\phi}\simeq1.4h^{2}\left(\frac{\Omega_{\phi0}h_{0}^{2}}{0.1}\right)^{1/2}\left(\frac{g_{\star F}}{10}\right)^{1/4}\mathrm{\left(\frac{x_{F}}{25}\right)^{-3/4}TeV}~.\label{Eq:WIMP IDM}
\end{equation}
where $g_{\star F}$ corresponds to the number of relativistic degrees of freedom at freeze-out and $x_{F}=m_{\phi}/T_{F}$, with $T_{F}$ being the freeze-out temperature. 

We can now relate equations \eqref{Eq:evap condition N 1} and \eqref{Eq:WIMP IDM} to obtain a restriction on
the right-handed neutrino mass in order to obtain the measured dark matter relic abundance:
\begin{equation}
m_{\phi}\gtrsim50\left(\frac{\Omega_{\phi0}h_{0}^{2}}{0.1}\right)^{1/2}\left(\frac{g_{\star F}}{10}\right)^{1/2}\left(\frac{x_{F}}{25}\right)^{-3/4}\left(\frac{M_{1}}{1\ \mathrm{GeV}}\right)^{1/2}\mathrm{\,keV}~.
\label{Eq:evap condition N scatt}
\end{equation}
Let us now consider evaporation through scatterings off the SM degrees of freedom, taking as example the scattering with the Higgs boson, represented in figure \ref{fig:Feyn1}. The corresponding loop diagram can be roughly estimated as yielding an effective four-point interaction with coupling  $g'\sim\frac{hy_{eff}}{\sqrt{16\pi^{2}}}$ up to $O(1)$ factors. The resulting cross section, for a center of mass energy $\sqrt{s}\sim T$, corresponds to 
\begin{equation}
\sigma_{\phi H}\simeq\frac{h^{2}y_{eff}^{2}}{256\pi^{3}T^{2}}~.
\label{Eq:Higgs scat cross}
\end{equation}
Using the number density of relativistic Higgs bosons, $n_{H}=\frac{2}{\pi^2}\zeta(3)T^{3}$, and that $\Gamma_{evap}\simeq n_{H}\left\langle \sigma v\right\rangle $, the scattering rate can be estimated as 
\begin{equation}
\Gamma_{evap}^{(h)}\simeq\frac{\zeta(3)}{128\pi^{5}}h^{2}y_{eff}^{2}T\,.\label{Eq:Decay width scattering w Higgs}
\end{equation}
Note that, at one-loop order, scattering with the left-handed neutrinos, present in the thermal bath, could also contribute to the evaporation process, as represented in figure \ref{fig:Feyn1}. However, the effective interaction term, $\phi^2\nu^2_{\ell}$, is a dimension-5 operator,  meaning that the effective coupling has an $M_1^{-1}$ suppression factor, leading to a $\sim T^2/M_1^2$ suppression of the cross section as compared to the Higgs scattering cross section. This leads to $\Gamma_{evap}^{(\nu)}\propto T^{3}/M_1$ and, as a result, the evaporation rate drops faster than the Hubble rate, having no impact on the evaporation process.    

In order to have an effective evaporation before electroweak symmetry breaking, such that relativistic Higgs particles are still present in the cosmic bath, we then obtain the condition:
\begin{equation}
\frac{\Gamma_{evap}^{(h)}}{H}\Bigl|_{T=100\ \mathrm{GeV}}=\frac{\zeta(3)}{128\sqrt{90}\pi^{6}}g_{\star}^{-1/2}\frac{M_{P}}{T}h^{2}y_{eff}^{2}\simeq10^{11}g_{\star}^{-1/2}h^{2}y_{eff}^{2}\gtrsim1~.\label{Eq:evap condition H 1}
\end{equation}
In the case $T_{R}>M_{1}$, evaporation of the inflaton condensate may thus proceed through scatterings with both the right-handed neutrinos and the Higgs boson, such that eqs \eqref{Eq:evap condition N 1} and \eqref{Eq:evap condition H 1} must be taken into account. If, however, $T_R <M_1$, at reheating right-handed neutrinos are already non-relativistic and have already decayed into SM particles as we have seen above. In this case evaporation of the condensate is dictated by inflaton scatterings off Higgs bosons, its efficiency being dictated by  \eqref{Eq:evap condition H 1}. 

In this discussion we have obtained relations and constraints on the $\nu$IDM model parameters. The model is defined by six parameters\footnote{There is an additional parameter $M_3$ in the case with normal light neutrino hierarchy, but here we will consider only the case where $N_1$ and $N_3$ give similar contributions to the non-zero mass eigenvalue $m_3$.} - the couplings $\xi$, $\lambda$, $h$ and $y_{eff}$, and in addition the inflaton and right-handed neutrino Majorana masses, $m_\phi$ and $M_1$, respectively. The three relations in equations \eqref{Eq:nmc and lambda}, \eqref{Eq:WIMP IDM} and \eqref{Eq:M1} then reduce the number of independent parameters to three. These relations then simplify the kinematical and evaporation constraints in equations \eqref{Eq:M1 bounds IDM}, \eqref{Eq:evap condition N scatt} and \eqref{Eq:evap condition H 1}, which can be written as:
\begin{align}
\frac{m_{\phi}}{2} & <M_{1}<\frac{3\sqrt{3}\ \lambda^{\frac{1}{2}}M_{p}m_{\phi}}{4\pi^{2}\ \mathrm{TeV}}\label{Eq:decay and reheating bound}\ ,\\
m_{\phi}\gtrsim & 4.9\times10^{-5}\left(\frac{M_{1}}{1\ \mathrm{GeV}}\right)^{1/2}\mathrm{GeV}\label{Eq:evap bound N scatt}\ ,\\
m_{\phi}\gtrsim & 1.7\times10^{8}\left(\frac{\mathrm{GeV}}{M_{1}}\right)\mathrm{GeV}.\label{Eq:evap bound H scatt}\
\end{align}
where we have also taken into account the kinematical condition for the incomplete decay of the inflaton field discussed above, $M_1> m_\phi/2$. 

\begin{figure}[H]
	\centering
	\includegraphics[totalheight=7.5cm]{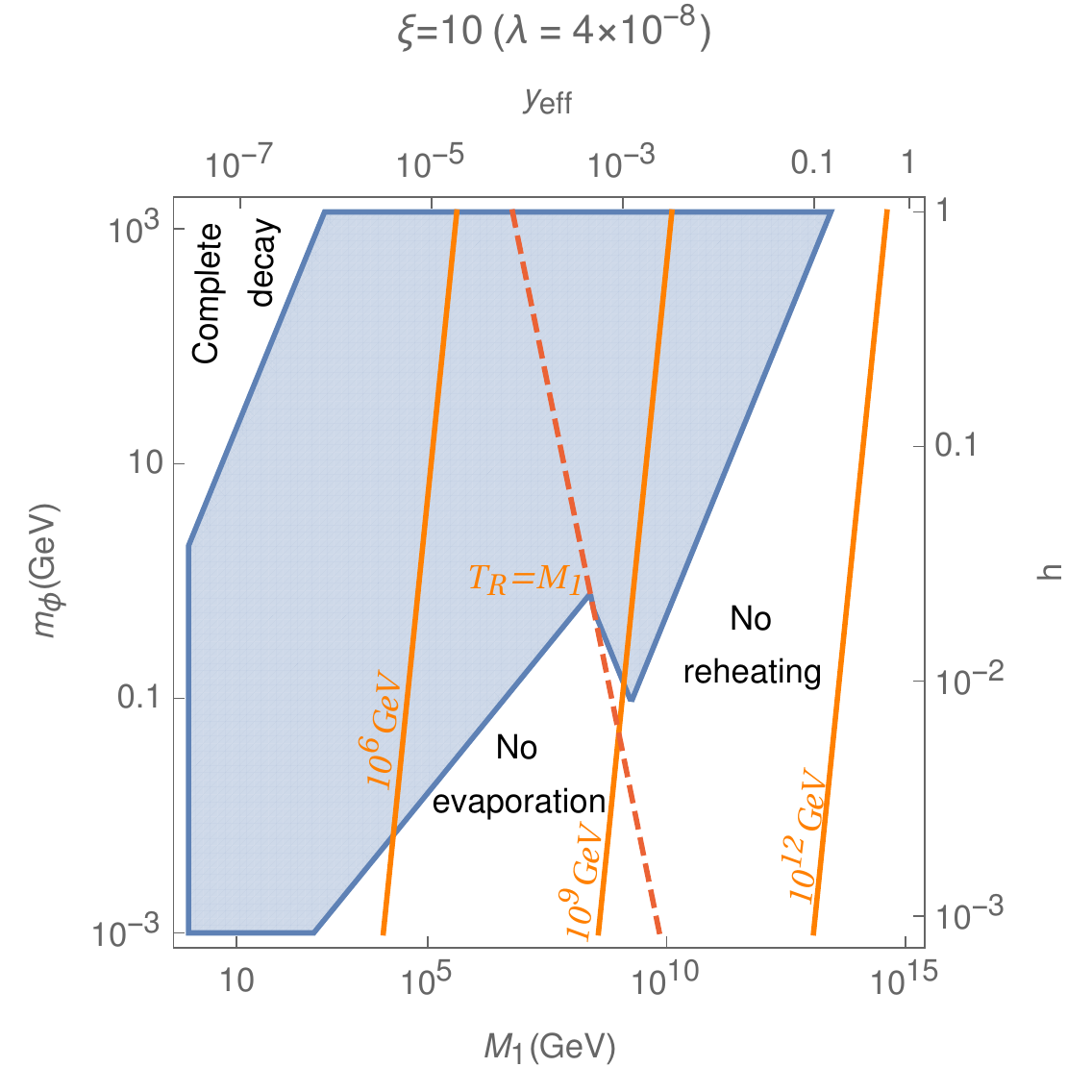}
	\includegraphics[totalheight=7.5cm]{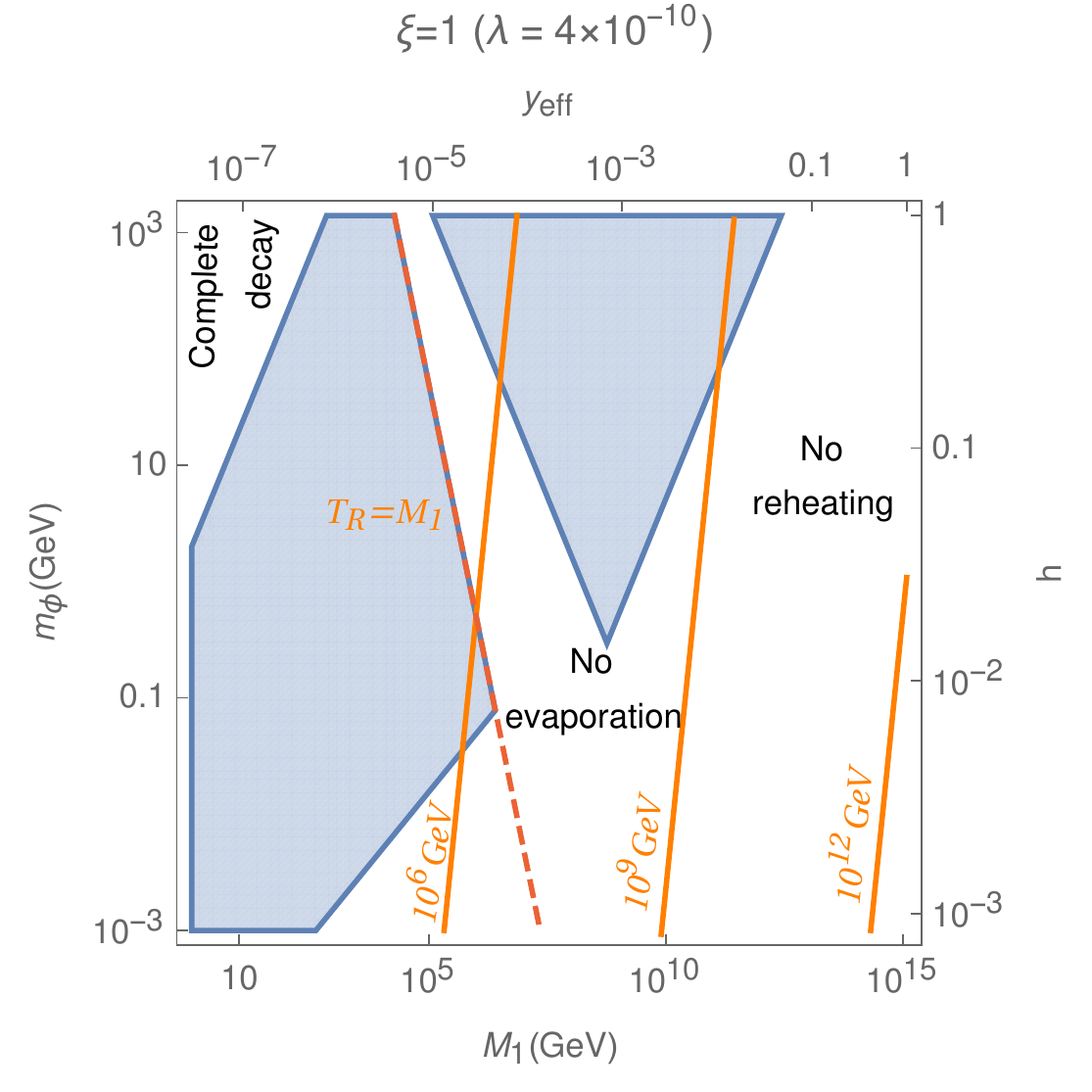}
	\includegraphics[totalheight=7.5cm]{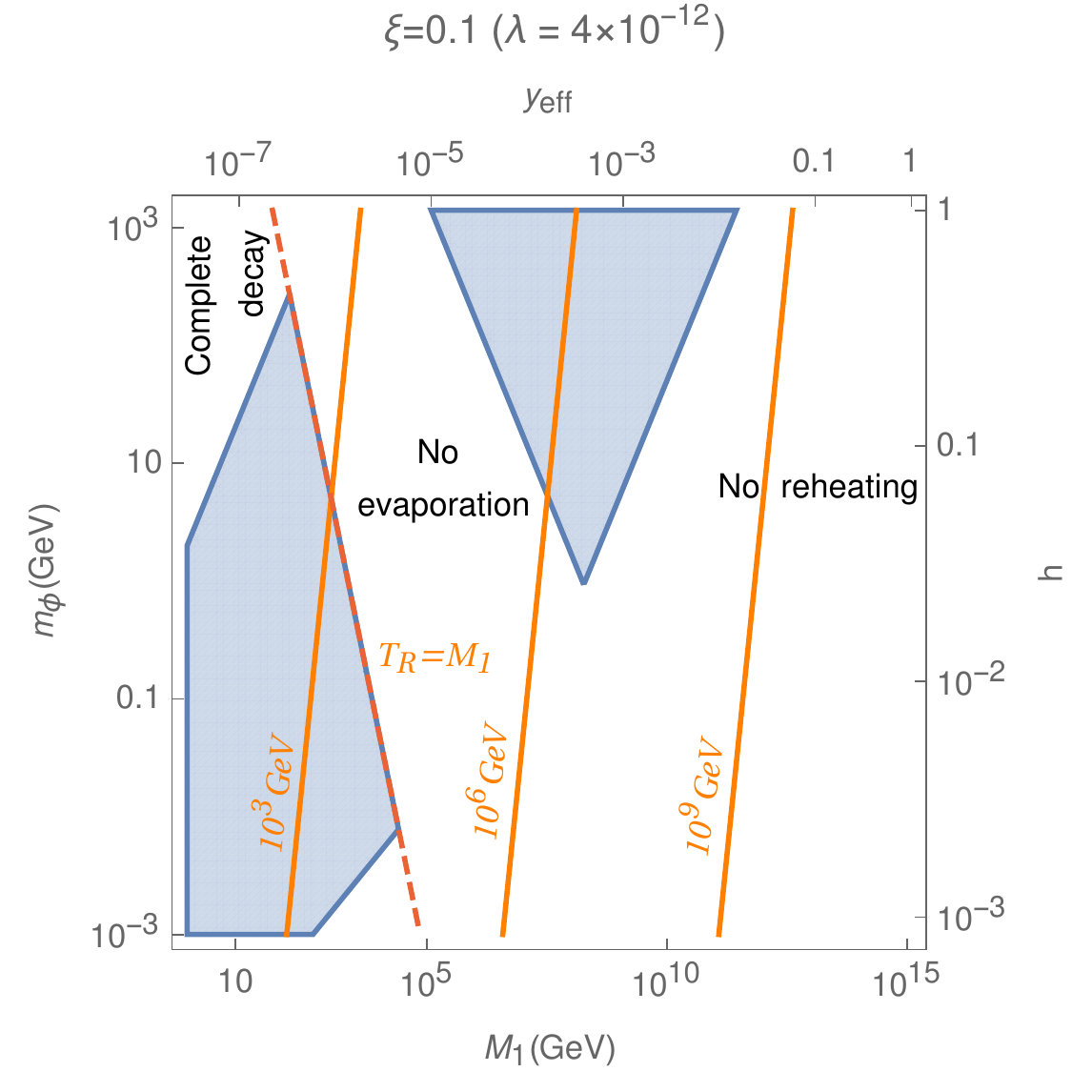}
	
	\caption{Constraints on model parameters for $\xi=0.1,\,1,$ and $10$, taking $g_\star =100$. The shaded blue regions correspond to parameter values satisfying all constraints in the WIMPlaton scenario. Orange lines represent different values for the reheating temperatures and the dashed line, $M_{1}=T_{R}$, limits the evaporation conditions. Condition \eqref{Eq:decay and reheating bound} defines the ``complete decay" and ``no reheating" regions and conditions \eqref{Eq:evap bound N scatt},\eqref{Eq:evap bound H scatt} limit the ``no evaporation" region. 
	} \label{fig:IDM}
\end{figure}

Taking into account that all couplings should lie in a regime where perturbation theory is valid, we represent these constraints in figure \ref{fig:IDM} for fixed values of $\xi\,(\lambda)$.

As one can see in this figure, reheating temperatures are always well above the BBN temperature. We have thus obtained a broad range of parameters where all of the above mentioned constraints are satisfied. On the one hand, scenarios with large values of the neutrino Majorana mass generically imply larger values of the inflaton mass, with $m_\phi \lesssim$1 TeV, as well as larger Yukawa couplings $h$ and $y_{eff}$. These are also the scenarios with the largest values of the reheating temperature. On the other hand, we may also have consistent WIMPlaton models with sub-GeV inflaton and Majorana masses, although these typically require very small values of the neutrino effective Yukawa coupling.

Although there are scenarios with sufficiently small values of the inflaton Yukawa coupling, $h$, within the allowed parametric regions in figure \ref{fig:IDM}, there are also regimes with large Yukawa couplings, up to $h\sim 1$. Since these may lead to significant radiative corrections to the inflaton self-coupling $\lambda$, one may worry that large Yukawa couplings are incompatible with the parametrically small values of $\lambda$ required by the amplitude of the primordial spectrum of curvature perturbations, which depends on the value of the non-minimal coupling $\xi$ as discussed earlier. We may, of course, accept some degree of fine-tuning in the model, but an alternative (and arguably more natural) possibility would be to consider supersymmetric versions of the $\nu$IDM model, where radiative corrections from the right-handed neutrinos $N_1$ and $N_2$ can be partially cancelled by those of their scalar superpartners. Although a detailed supersymmetric extension of the model is outside the scope of this work, we note that the addition of right-handed sneutrinos, which are charged under the discrete interchange symmetry, will only make the inflaton decay more efficiently, guaranteeing nevertheless its stability at late times and thus preserving the fundamental features of the $\nu$IDM model.

The analysis above also shows that there are regions in parameter space where the evaporation conditions are not satisfied, but for which nevertheless one may obtain an efficient reheating of the Universe along with an incomplete decay of the inflaton field that leaves a stable remnant. This thus motivates looking for scenarios where condensate evaporation is inefficient and the inflaton remains as an oscillating scalar field in all the post-inflationary cosmic history, as we discuss below.

\subsection{Oscillating scalar field dark matter scenario}
\label{sub4.4}

As we have discussed above, the inflaton decays until this becomes kinematically blocked for field amplitudes below:
\begin{equation}
\phi_{DR}=\frac{M_{1}}{h}~.\label{Eq: phiDR}
\end{equation}
If evaporation processes are inefficient, while the quartic term dominates the inflaton potential the condensate will behave as a dark radiation (DR) fluid. This lasts until the quadratic term becomes dominant, when the oscillation amplitude becomes smaller than
\begin{equation}
\phi_{CDM}=\sqrt{\frac{1}{2}\frac{m_{\phi}^{2}}{\lambda}}~,\label{Eq:phi cdm}
\end{equation}
after which the oscillating inflaton field starts behaving as cold dark matter (CDM). At this stage, the ratio $n_\phi/s$ becomes constant until the present day, and we can relate it to the measured dark matter density:
\begin{equation}
\Omega_{\phi0}=\frac{\rho_{\phi0}}{\rho_{c0}}=\frac{m_{\phi}n_{\phi0}}{3H_{0}^{2}M_{P}^{2}}=\frac{m_{\phi}s_{0}}{3H_{0}^{2}M_{P}^{2}}\left(\frac{n_{\phi}}{s}\right)_{CDM}.\label{Eq:energy ratio}
\end{equation}
Using that, between $\phi_{DR}$ and $\phi_{CDM}$, the inflaton field behaves as a radiation fluid and that $T_{DR}\simeq T_{R}$, we have:
\begin{equation}
\frac{\phi_{DR}}{\phi_{CDM}}\simeq\frac{T_{R}}{T_{CDM}}~.\label{Eq: Tcdm}
\end{equation}
Now, the inflaton number density-to-entropy ratio is given by:
\begin{equation}
\left(\frac{n_{\phi}}{s}\right)_{CDM}=\left(\frac{\rho_{\phi}}{m_{\phi}s}\right)_{CDM}=\frac{\frac{1}{2}m_{\phi}\phi^2_{CDM}}{\frac{2\pi^{2}}{45}g_{\star CDM}T_{CDM}^{3}}~.\label{Eq:n e s ratio}
\end{equation}
Using Eqs.~\eqref{Eq:phi^4 reheating T}, \eqref{Eq: Tcdm}, \eqref{Eq:n e s ratio},  and $s_{0}=\frac{2\pi^{2}}{45}g_{\star0}T_{0}^{3}$ in Eq.~\eqref{Eq:energy ratio}, we then obtain
\begin{equation}
\Omega_{\phi0}=\frac{\left(\frac{2}{5}\right)^{3/4}\pi^{7/2}}{27\times3^{1/4}}\left(\frac{g_{\star R}^{3/4}g_{\star0}}{g_{\star CDM}}\right)\frac{T_{0}^{3}m_{\phi}M_{1}}{\lambda^{3/4}h^{2}H_{0}^{2}M_{P}^{3}}~,
\end{equation}
and for $H_{0}=10^{-42}h_{0}\ \mathrm{GeV}$, $g_{\star0}=3.91$ and $T_{0}=2.4\times10^{-13}\ \mathrm{GeV}$ this yields:
\begin{equation}
\Omega_{\phi0}h_{0}^{2}\simeq 3\times10^{-9}\left(\frac{g_{\star R}^{3/4}}{g_{\star CDM}}\right)\frac{1}{\lambda^{3/4}h^{2}}\left(\frac{m_{\phi}M_{1}}{\mathrm{GeV}^{2}}\right)~.
\end{equation}
The inflaton mass yielding the measured dark matter abundance in this oscillating scalar field scenario is thus given by:
\begin{equation}
m_{\phi}\simeq 3.2g_{\star CDM}g_{\star R}^{-3/4}h^2 \xi^{3/2}\left(\frac{\Omega_{\phi0}h_{0}^{2}}{0.1}\right)\left(\frac{\mathrm{GeV}}{M_{1}}\right)\ \mathrm{GeV}
\label{Eq:mphi noevap}.
\end{equation}
Similarly to the previous subsection, we can study the allowed parameter space using the relations \eqref{Eq:nmc and lambda}, \eqref{Eq:M1} and \eqref{Eq:mphi noevap}, imposing the kinematical bound \eqref{Eq:M1 bounds IDM} and the reciprocal of the efficient evaporation conditions \eqref{Eq:evap condition N scatt} and \eqref{Eq:evap condition H 1}, such that the inflaton remnant survives as an oscillating scalar field. We then find:
\begin{align}
\frac{m_{\phi}}{2} & <M_{1}<\frac{7.7\times 10^{10}\, M_{1}m_{\phi}}{\lambda^{1/4}\ \mathrm{GeV}},\label{Eq:decay and reheating bound2}\\
m_{\phi}\lesssim & \,0.14\left(\frac{\mathrm{GeV}}{M_{1}}\right)^{1/2}\lambda^{3/4}\ \mathrm{GeV},\label{Eq:noevap bound N scatt}\\
m_{\phi}\lesssim & \, 4.9\times10^{11}\left(\frac{\mathrm{GeV}}{M_{1}}\right)^{2}\lambda^{3/4}\ \mathrm{GeV}~.\label{Eq:noevap bound H scatt}
\end{align}
In this case, we impose in addition that all couplings, in particular the inflaton Yukawa coupling, $h$, remain within the limits of a perturbative analysis, and that $ T_{CDM}>T_{eq}$ and $T_{R} > T_{BBN}$, where $T_{eq}=0.79\ \mathrm{eV}$ is the temperature at matter-radiation equality and $T_{BBN}\simeq10\ \mathrm{MeV}$. 

\begin{figure}[htbp]
	\centering
	\includegraphics[totalheight=8.3cm]{{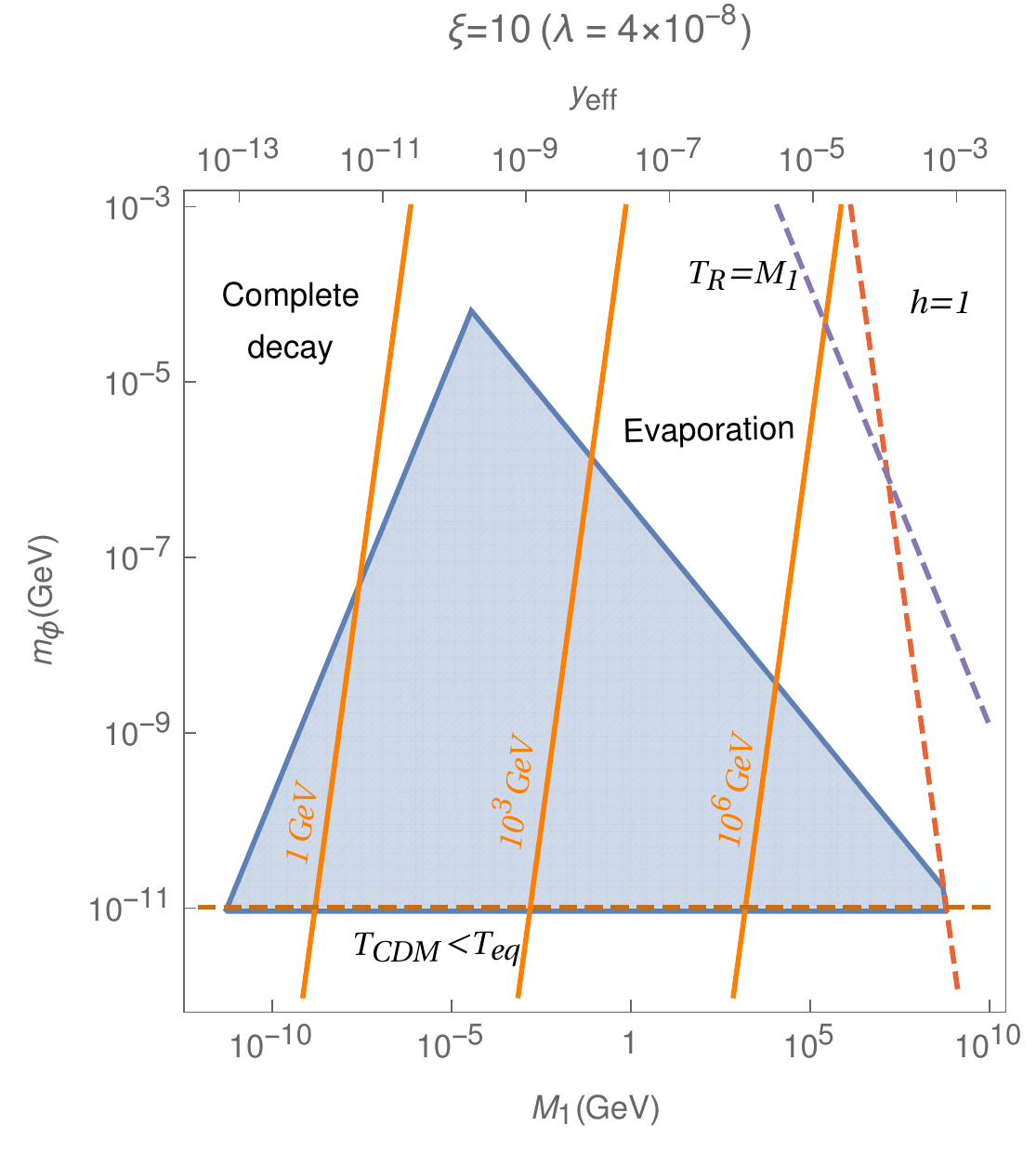}}
	\includegraphics[totalheight=8.3cm]{{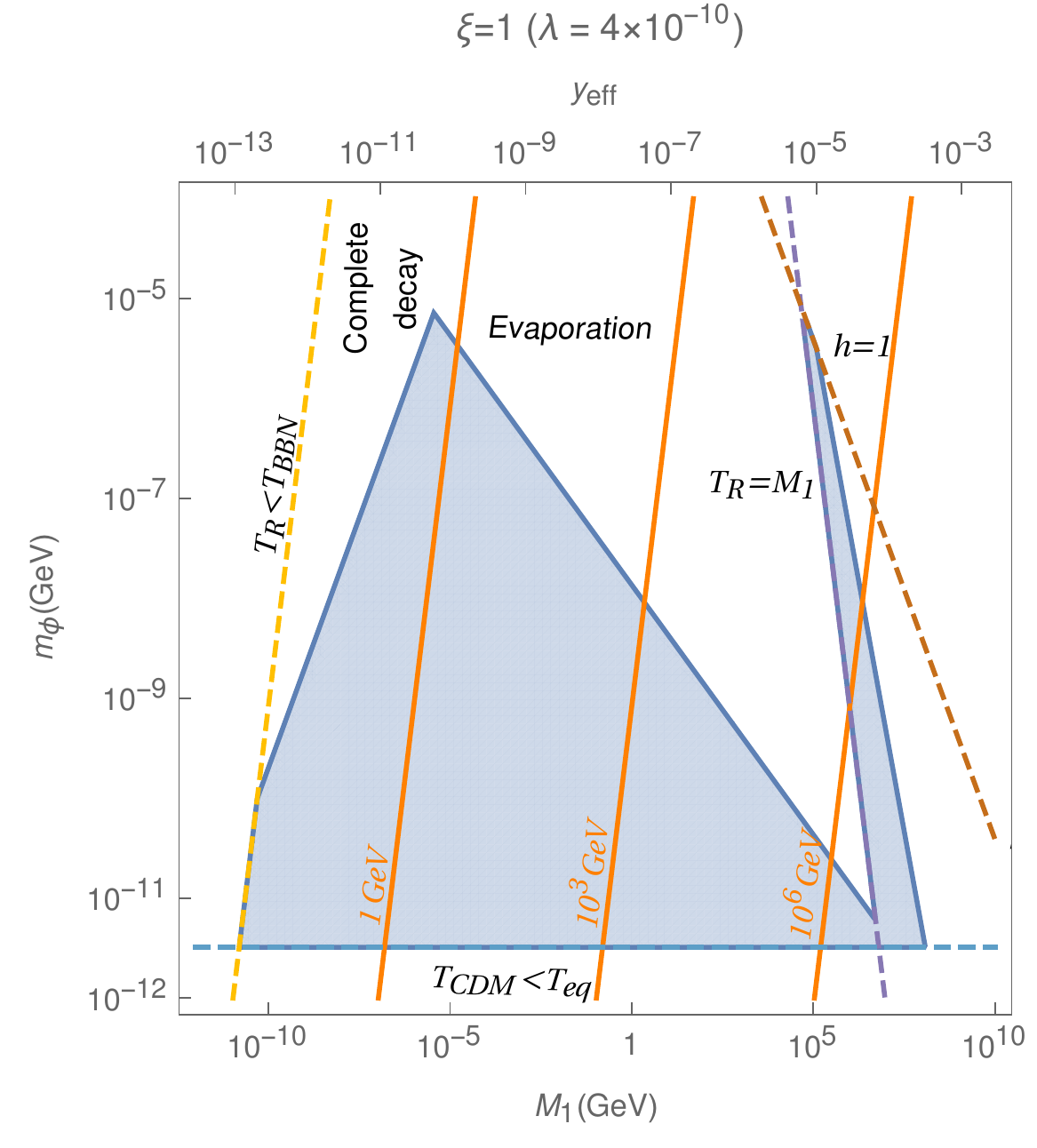}}\vspace{0.5cm}
	\includegraphics[totalheight=8.3cm]{{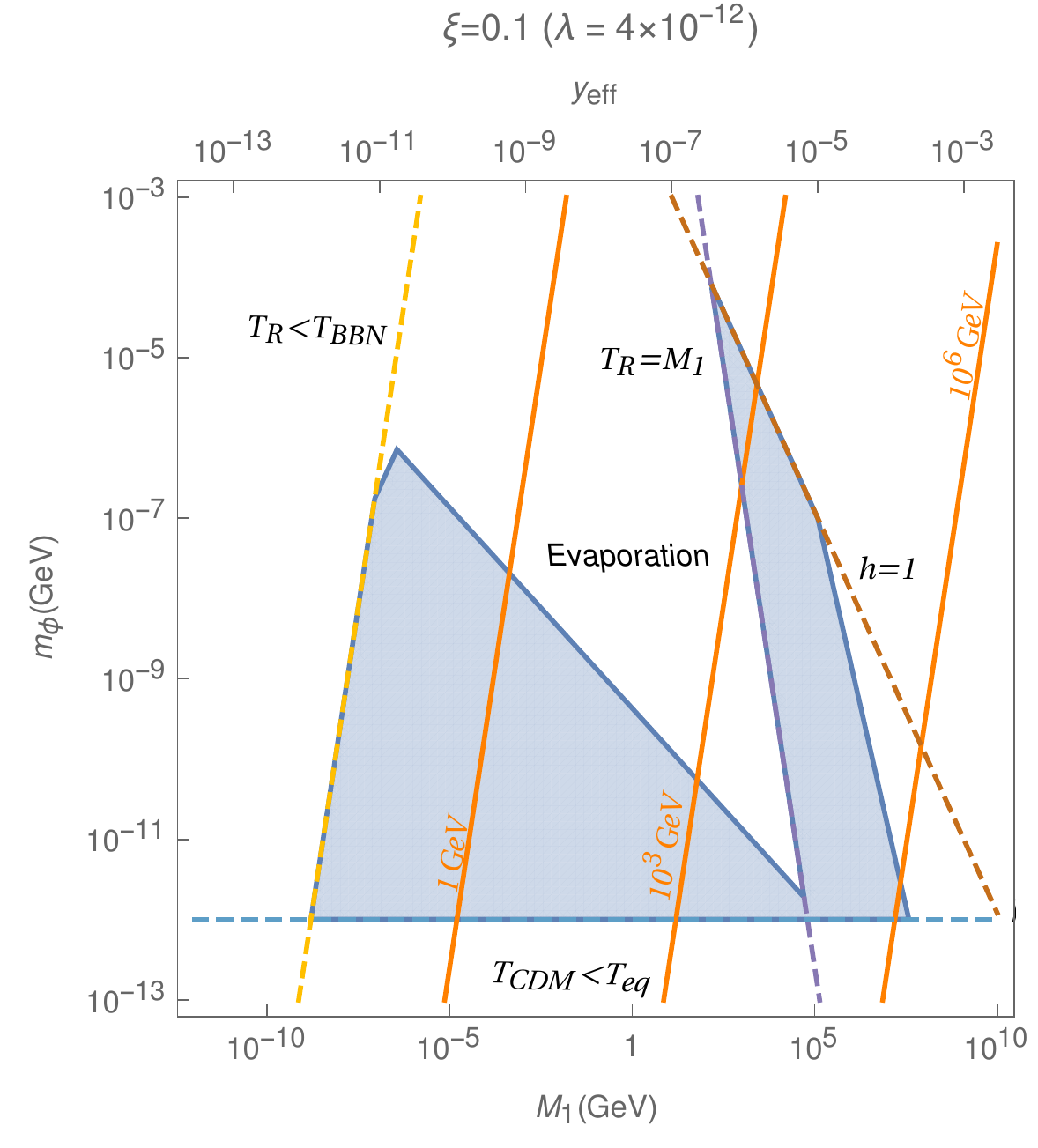}}
	
	\caption{Constraints on model parameters for $\xi=0.1,\,1,$ and $10$, taking $g_\star =100$. The shaded blue regions correspond to parameter values satisfying all constraints in the oscillating scalar field dark matter scenario. Orange lines represent different values for the reheating temperature and the blue dashed line, $M_{1}=T_{R}$, limits the evaporation conditions. The $h=1$, $T_R<T_{BBN}$ and $T_{CDM}<T_{eq}$ limits are also represented by dashed lines. The condition \eqref{Eq:decay and reheating bound2} defines the ``complete decay" and ``no reheating" regions and conditions \eqref{Eq:noevap bound N scatt} and \eqref{Eq:noevap bound H scatt} limit the ``evaporation" region. 
	} \label{fig:Noevap}
\end{figure}

We represent the resulting parameter space in figure \ref{fig:Noevap}, where one can see that the conditions for avoiding the evaporation of the inflaton condensate generically lead to smaller values of the inflaton mass, in the sub-GeV range, as well as lower values of the Majorana mass and Yukawa couplings. This is, in fact, typical of dark matter candidates that never thermalize with the cosmic plasma, as other oscillating scalar fields such as axions (see e.g.~\cite{Marsh:2015xka}) or Higgs-portal scalar field models \cite{Bertolami:2016ywc, Cosme:2017cxk, Cosme:2018nly}. The resulting reheating temperatures are also much lower than for the WIMPlaton scenario, since the smaller values of the couplings make the reheating process much less efficient.

\section{Thermal leptogenesis in the $\nu$IDM model}
\label{sec5}

One of the most important consequences of the seesaw mechanism, introduced to account for the observed lightness of neutrino masses, is a lepton number violation. At the same time, this generation of a neutrino mass generally requires new CP violating phases in Yukawa interactions and new heavy singlet neutrinos that may decay out of equilibrium. Thus, all
three Sakharov conditions seem to be satisfied within such a mechanism  \cite{Sakharov:1967dj}. The challenge is then to assess whether such a scenario can actually account for the observed baryon asymmetry. 
 
In this section, we will embed a thermal leptogenesis scenario within the $\nu$IDM model, via a thermal production of the third right-handed neutrino, $N_{3}$, assuming it is lighter than the ones produced by the inflaton decay, $M_{3}\lesssim M_{1}$. A lepton asymmetry may then be generated in $N_3$ decays, and this may later be converted into a baryon asymmetry by the $B+L$ violating electroweak sphaleron processes (see e.g.~\cite{Rubakov:1996vz}). We will restrict our analysis to the WIMPlaton scenario, where sufficiently high reheating temperatures for thermal leptogenesis can be attained, as we discuss below. Here we will describe only the basic features of such a leptogenesis scenario, and the reader is referred to the literature for a more detailed description  \cite{Davidson:2008bu,Nir:2007zq}.

The baryon asymmetry resulting from the decay of an initial thermal population of relativistic right-handed neutrinos $N_3$ and subsequent sphaleron processing is given by  \cite{Davidson:2008bu, Nir:2007zq}: 
\begin{equation}
Y_{\Delta B}\simeq\left(\frac{n_{N_{3}}}{s_{T\gg M_{3}}}\right) C_{sphal}\eta\epsilon=\left(\frac{135\zeta(3)}{4\pi^{4}g_{\star}}\right) C_{sphal}\eta\epsilon~,\label{Eq:Baryon asymmetry thermal L}
\end{equation}
where $C_{sphal}$ represents the dilution of the lepton asymmetry by electroweak sphaleron processes, $0<\eta<1$ parametrizes the efficiency of the lepton number violating interactions, taking into account washout processes such as inverse decays, and $\epsilon$ quantifies the amount of CP-violation in $N_3$ decays. The latter is given by:
\begin{equation}
\epsilon\equiv\frac{\Gamma\left(N_{3}\rightarrow HL\right)-\Gamma\left(N_{3}\rightarrow\bar{H}\bar{L}\right)}{\Gamma\left(N_{3}\rightarrow HL\right)+\Gamma\left(N_{3}\rightarrow\bar{H}\bar{L}\right)}~,
\end{equation}
where $L$ and $H$ represent the lepton and Higgs doublets and $N_{3}=\bar{N}_{3}$, since the heavy neutrino is a Majorana particle. 

This asymmetry results from the interference of tree-level and one-loop amplitudes \cite{Davidson:2008bu} corresponding to the following Feynman diagrams:
\begin{figure}[H]
	\centering
	
	\begin{tikzpicture}[line width=1 pt, scale=1.3]
	\draw[scalar] (0:0)--(-40:1);
	\draw[fermion] (0:0)--(40:1);
	\draw[fermionnoarrow] (180:1)--(0:0);
	\node at (-40:1.2) {$H$};
	\node at (40:1.2) {$L_{\ell}$};
	\node at (180:1.2) {$N_{3}$};
	\begin{scope}[shift={(-0.5,0)}]
	\draw (125:.15) -- (-55:.15);
	\draw (55:.15) -- (-125:.15);			
	\end{scope}	
	\end{tikzpicture}
	\begin{tikzpicture}[line width=1 pt, scale=1.3]
	\draw[fermion] (-1,0)--(0,0);
	\node at (-0.5,0.3) {$N_{3}$};
	\draw[fermionbar] (1,0) arc (0:180:.5);
	\draw[scalar] (1,0) arc (0:-180:.5);
	\node at (0.5,0.8) {$L_{\ell}$};
	\node at (0.5,-0.8) {$H$};
	\draw[fermionnoarrow] (1,0) --(2,0);
	\node at (1.5,0.3) {$N_{1,2}$};
	\begin{scope}[shift={(2,0)}]
	\draw[fermion] (-40:1)--(0,0);
	\draw[scalar] (0:0)--(40:1);
	\node at (-40:1.2) {$L_{\ell}$};
	\node at (40:1.2) {$H$};	
	\end{scope}
	\begin{scope}[shift={(1.5,0)}]
	\draw (125:.15) -- (-55:.15);
	\draw (55:.15) -- (-125:.15);			
	\end{scope}		
	\end{tikzpicture}
	\begin{tikzpicture}[line width=1 pt, scale=1.3]
	\draw[scalar] (0:0)--(-40:1);
	\draw[fermion] (0:0)--(40:1);
	\draw[fermion] (180:1)--(0:0);
	\node at (-60:0.8) {$H$};
	\node at (60:0.8) {$L_{\ell}$};
	\node at (180:1.2) {$N_{3}$};
	\draw[fermionnoarrow] (-40:1)--(40:1);
	\draw[fermion] (-40:1)--(2,-0.64);
	\draw[scalar] (40:1)--(2,0.64);
	\node at (2.2,-0.64) {$L_{\ell}$};
	\node at (2.2,0.64) {$H$};
	\node at (1.1,0) {$N$};
	\begin{scope}[shift={(0.766044443,0)}]
	\draw (125:.15) -- (-55:.15);
	\draw (55:.15) -- (-125:.15);			
	\end{scope}		
	\end{tikzpicture}
	\caption{Feynman diagrams contributing to the CP asymmetry $\epsilon $.} \label{fig:Feyn2}
\end{figure}
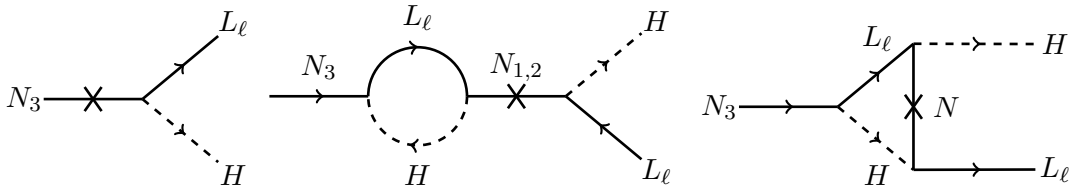

The value of the CP-asymmetry is bounded from above, as originally shown by Davidson and Ibarra \cite{Davidson:2002qv} (see also e.g.~\cite{Davidson:2008bu, Nir:2007zq}):
\begin{equation}
\left|\epsilon\right|\leq\frac{3M_{3}m_{\mathrm{max}}}{8\pi v^{2}}~,
\label{Eq:Davidson_ibarra bound}
\end{equation}
where $m_{\mathrm{max}}$ denotes the largest light neutrino mass eigenvalue. From here one may obtain a lower bound on $M_{3}$ and consequently on the required reheating temperature as we discuss below. 

The non-equilibrium condition necessary for leptogenesis is ensured by the expansion of the Universe. With the decrease of the Hubble parameter, after reheating and while in a radiation dominated Universe, interactions rates will change from faster to slower than $H$ \cite{Davidson:2008bu} so that $N_3$ may decay out-of-equilibrium. 

As we have seen above, in the $\nu$IDM model the radiation bath after inflaton-radiation equality may include the right-handed neutrinos $N_1$ and $N_2$ resulting from inflaton decay, as well as their decay products if $T_R\lesssim 10 g_\star^{-1/4} M_1$. If the reheating temperature exceeds the mass of the third and lightest right-handed neutrino, $T_R>M_3$ and if the latter is produced faster than expansion, $\Gamma_{prod}>H$, a thermal population of $N_{3}$ particles may be generated. Since there is no direct decay from the inflaton into the third right handed neutrino, its production channels are trough inverse decays and in 2-2 scatterings involving the electroweak gauge bosons and the top quark. We can estimate such a rate as \cite{Davidson:2008bu}
\begin{equation}
\Gamma_{prod}\sim\sum_{\ell}\frac{y_{t}^{2}\left|y_{3\ell}\right|^{2}}{4\pi}T~.
\end{equation}
Since $y_{t}\sim1$, this exceeds the $N_3$ decay width $\Gamma_{D}=\left[y_{\nu}y_{\nu}^{\dagger}\right]_{33}M_{3}/(8\pi)$ for $T>M_3$. We may then ensure that a thermal population of $N_3$ is produced for $\Gamma_D> H$. In the radiation era, we have
\begin{equation}
\frac{\Gamma_D}{H}\simeq0.2\,g_{\star}^{-1/2}\,y_{3\,eff}^2\frac{ M_P}{T}~,
\end{equation}
where $y_{3\,eff}^2=\sum_{\ell} \left| y_{3\ell}\right| ^2$. Imposing that $\Gamma_D\gtrsim H$ at temperatures $T\gtrsim 10 M_3$, we obtain the following lower bound on the effective $N_3$ Yukawa coupling:
\begin{equation}
y_{3\,eff}^2\gtrsim1.7\times10^{-16}\frac{M_3}{\mathrm{GeV}}~.\label{Eq:bound y_3eff}
\end{equation}
Using the results obtained in section \ref{sec3} for the light neutrino mass spectrum, namely Eqs.~\eqref{Eq:solution 1 n mass IDM} and \eqref{Eq:solution 2 n mass IDM}, and the experimental values for the neutrino squared mass differences, we have 
\begin{equation}
y_{3\,eff}^2\simeq1.6\times10^{-15} \frac{M_3}{\mathrm{GeV}}~,
\end{equation} 
for an inverted mass hierarchy, and a similar result should apply up to $\mathcal{O}(1)$ factors for a normal hierarchy, provided that the contributions of $N_{1,2}$ and $N_3$ to the non-zero mass eigenvalue $m_3$ are comparable. We thus conclude that the light neutrino mass spectrum is consistent with the thermal production of an $N_3$ population without imposing additional constraints on the model.

Interactions that are faster than $H$ impose chemical and kinetic equilibrium conditions on the involved particles and in this scenario
the total lepton asymmetry $Y_{L}\simeq0$. Although a lepton asymmetry may arise in $N_{3}$ decays, any asymmetry is washed out by processes like inverse decays or scatterings between the leptons and the Higgs or the leptons and $N_{3}$.

An asymmetry will only arise once, for instance, inverse decays become inefficient:
\begin{equation}
\Gamma_{ID}(HL\rightarrow N_{3})\simeq\Gamma_{D}e^{-M_{3}/T_{*}}<H\,.
\end{equation}
At $T_{*}$ the inverse decay rate becomes smaller than $H$, blocking the production of $N_{3}$, and the remaining $N_{3}$ density becomes Boltzmann suppressed, making these right-handed neutrinos decay out-of-equilibrium. 

The $N_{3}$ decay violates lepton number, since no consistent attribution of a lepton number can be assigned to the right-handed neutrinos with the presence of both Majorana and Yukawa mass terms. The heavy neutrino mass eigenstate, being a Majorana particle and hence its own anti-particle, can decay into both $LH$ and $\bar{L}\bar{H}$. With an asymmetry in the decay rates, a net lepton asymmetry will be generated. The conversion to a baryon asymmetry is obtained through $(B+L)$-violating non-perturbative processes existent in the SM. 

The renormalizable SM Lagrangian conserves baryon number and the three lepton flavour numbers $L_{\ell}$. However, due to the chiral anomaly, there are non-perturbative processes that violate the combined total lepton number and baryon number, $B+L$. At zero temperature such gauge field configurations are called instantons and have highly suppressed rates, resulting in negligible $B+L$ violations \cite{Davidson:2008bu,Coleman:1978ae}. In the leptogenesis picture, where temperatures are well above the electroweak phase transition, the dominant $(B+L)$-violating processes are called sphalerons and have an approximate rate \cite{Burnier:2005hp} 
\begin{equation}
\Gamma_{B+L\,violation}\simeq250\alpha_{W}^{5}T~.
\end{equation}
Between the temperatures of the electroweak phase transition and $10^{12}\ \mathrm{GeV}$ this rate is faster than Hubble expansion. Thus, the lepton asymmetry produced in the $N_{3}$ decay, or the $B-L$ asymmetry, will be converted into a final baryon asymmetry following
\begin{equation}
Y_{\Delta B}\simeq\frac{12}{37}Y_{\Delta(B-L)}~,\label{Eq:baryon and lepton asymmetry ratio}
\end{equation}
where $Y_{\Delta(B-L)}$ is the $B-L$ asymmetry divided by the entropy density. 

\begin{figure}[h]
	\centering
	\includegraphics[totalheight=7.4cm]{{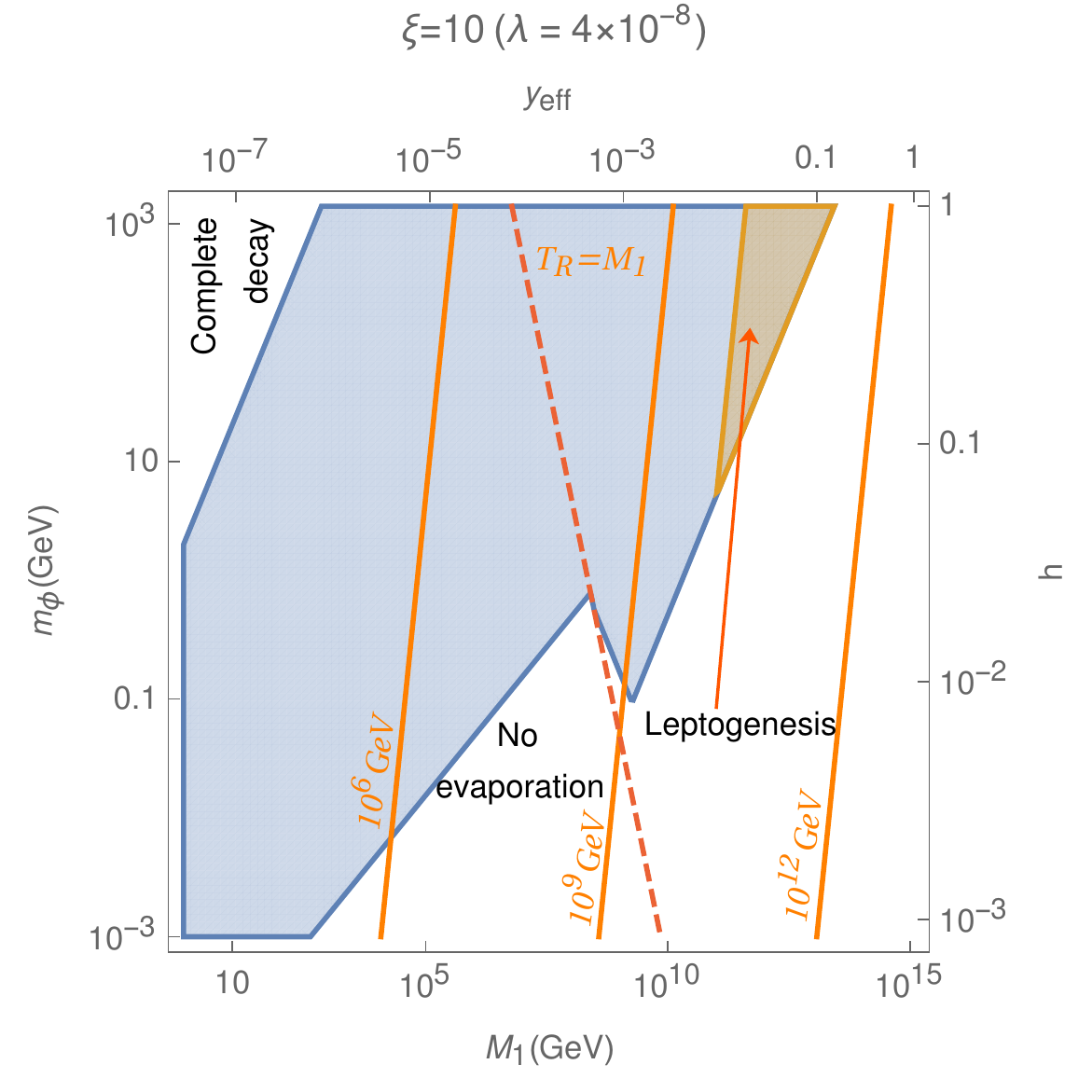}}
	\includegraphics[totalheight=7.4cm]{{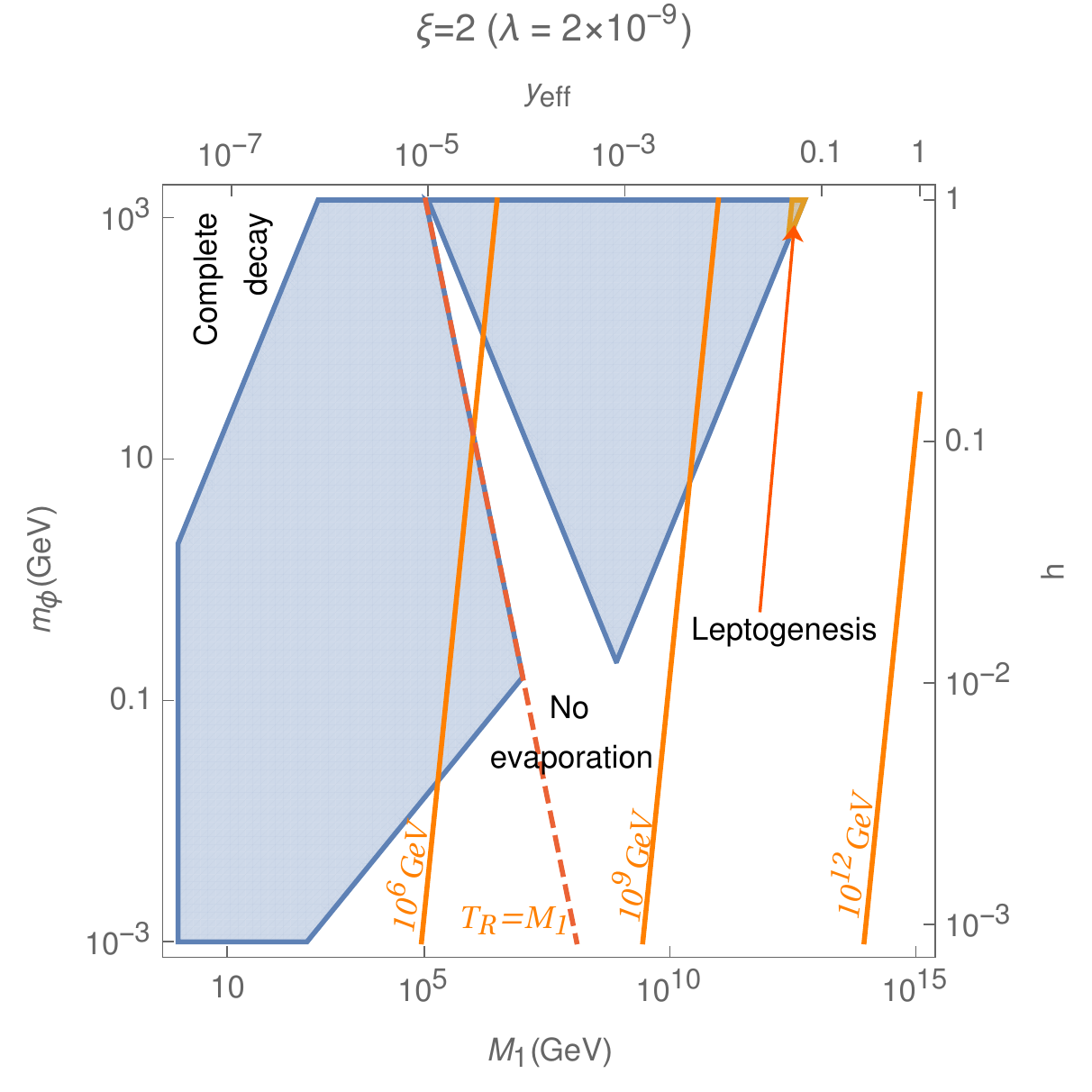}}
	\caption{Constraints on model parameters for $\xi=2,10$, taking $g_\star =100$. The shaded blue regions correspond to parameter values satisfying all constraints in the WIMPlaton scenario and the shaded orange regions correspond to consistent scenarios for thermal leptogenesis with $T_R> 10^{10}$ GeV. Orange lines represent different values for the reheating temperatures and the dashed line, $M_{1}=T_{R}$, limits the evaporation conditions. Condition \eqref{Eq:decay and reheating bound} defines the no decay and no reheating regions and conditions \eqref{Eq:evap bound N scatt},\eqref{Eq:evap bound H scatt} limit the no-evaporation region. 	
	} \label{fig:LIDM}
\end{figure}

Putting everything together, with equations \eqref{Eq:Baryon asymmetry thermal L} and \eqref{Eq:baryon and lepton asymmetry ratio},
we have
\begin{equation}
Y_{\Delta B}\simeq10^{-3}\epsilon\eta\,.
\end{equation}
Washout processes have been shown to result in an efficieny $0.01\lesssim\eta\lesssim0.1$ \cite{Davidson:2008bu}, and thus to account for the observed $Y_{\Delta B}\sim10^{-10}$ we need $\left|\epsilon\right|\gtrsim10^{-5}-10^{-6}$. The Davidson-Ibarra bound \eqref{Eq:Davidson_ibarra bound} and the light neutrino mass spectrum of the $\nu$IDM model then impose a lower bound on the $N_3$ Majorana mass:
\begin{equation}
M_{3}\gtrsim\frac{8\pi}{3}\frac{v^{2}}{m_{max}}\left|\epsilon\right|\simeq10^{9}\ \mathrm{GeV}~.
\end{equation}
Hence, to ensure that a population of relativistic $N_3$ right-handed neutrinos can be produced at reheating, we must require $T_R\gtrsim 10M_3$, i.e.
\begin{equation}
T_{R}\gtrsim10^{10}\ \mathrm{GeV}~.
\end{equation}
In figure \ref{fig:LIDM} we show the regions in parameter space where sufficiently large values of the reheating temperature can be obtained within the $\nu$IDM model, in the WIMPlaton scenario where the inflaton condensate evaporates.

As one can see, thermal leptogenesis is only realizable for $\xi\gtrsim2$ and in the region where ${T_{R}<M_{1}}$, due to the high reheating temperature required. Moreover, thermal leptogenesis can only occur in the upper limits for the parameters $h$, $m_{\phi}$ and $y_{eff}$, excluding parameters in the ``no-evaporation" regime. 

We note that this analysis assumes that condensate evaporation occurred after inflaton-radiation equality, but as we have mentioned above larger values of the reheating temperature can be attained if evaporation precedes reheating, thus possibly enlarging the available parameter space for thermal leptogenesis within the $\nu$IDM scenario.

Although we have considered only the case of thermal leptogenesis, other means of generating a lepton asymmetry could be available, for instance, through a non-thermal production of the right-handed neutrinos \cite{Davidson:2002qv,Asaka:2002zu}. In the $\nu$IDM model, the $N_1$ and $N_2$ right-handed neutrinos are, in fact, produced non-thermally by the decays of the inflaton field, so it would be interesting to investigate whether a lepton asymmetry can be generated in this case.

\section{Conclusions}
\label{sec6}

In this work, we have developed a unified model where we describe, using the same scalar field, both inflation and cold dark matter, due to an incomplete decay of the inflaton field into right-handed neutrinos. The introduction of right-handed neutrinos also allows for the generation of light neutrino masses through the seesaw mechanism and of the observed cosmological baryon asymmetry via thermal leptogenesis.

Based on the description of the incomplete decay of the inflaton originally proposed in \cite{Bastero-Gil:2015lga}, we have constructed a model where the inflaton can only decay into two of the three right-handed neutrinos, imposing a discrete interchange symmetry that forbids additional inflaton decay channels. We further imposed that inflaton decay is kinematically blocked at the minimum of the potential, for $\phi=0$, thus allowing for a stable inflaton remnant at late times. Nevertheless, oscillations of the right-handed neutrino masses due to Yukawa couplings to the inflaton field allow the latter to decay until its oscillation amplitude falls below a threshold value, thus providing a successful reheating of the Universe after inflation.

We have considered the simplest renormalizable form of the inflaton potential compatible with the discrete symmetry, including both a quadratic mass term and quartic self-interactions, as well as a non-minimal coupling to gravity that flattens out the potential at large field values. This allows for a successful ``plateau"-like period of slow-roll inflation compatible with the measured temperature and polarization CMB spectrum, according to Planck data, for a non-minimal coupling $\xi>0.008$, corresponding to an inflaton quartic self-coupling $\lambda>3\times 10^{-14}$.

Since inflation occurs for large $\phi$ values, the quartic potential dominates during the reheating period following inflation. By analyzing the dynamics of inflaton decays in this period, we have shown that a successful reheating, leading to a graceful-exit into the standard cosmological era, provided that the Majorana mass of the right-handed neutrinos $M_1\lesssim  6.6\times 10^{12} \xi h^2$ GeV, allowing for parametrically large values. We were also able to estimate analytically the reheating temperature, ensuring that it is parametrically above the required temperatures for BBN. The right-handed neutrinos may decay into SM degrees of freedom before or after reheating, depending on the parametric regime, and in general they decay before becoming non-relativistic.

Both the right-handed neutrinos $N_1$ and $N_2$, as well as their decay products, can scatter off low-momentum inflaton particles in the oscillating condensate and lead to its evaporation and thermalization. If this occurs, inflaton particles will later decouple and their abundance will freeze-out as standard WIMP-like candidates, which constitutes the ``WIMPlaton scenario". Otherwise, the inflaton remains as an oscillating condensate until the present day. 

We have analyzed both scenarios taking into account the conditions for incomplete decay, successful reheating and evaporation of the condensate, as well as the light neutrino mass spectrum and concluded that there is a broad range of inflaton and right-handed neutrino masses and couplings for which the inflaton can successfully account for all of the observed cold dark matter abundance. The WIMPlaton scenario typically involves larger mass scales and couplings, as well as larger values of the reheating temperature, than the oscillating scalar field scenario, the latter requiring sub-GeV inflaton masses.

Furthermore, we have shown that thermal leptogenesis can be naturally incorporated within our unified inflaton-dark matter scenario, through the $L$- and $CP$-violating decays of the right-handed neutrino $N_3$, which although decoupled from the inflaton field due to the underlying interchange symmetry may nevertheless be thermally produced in the reheating process. This implies reheating temperatures $T_R\gtrsim 10^{10}$ GeV, which can only be attained in the WIMPlaton scenario and for values of the non-minimal coupling to gravity $\xi\gtrsim 2$.

From the low energy perspective, the $\nu$IDM model makes a very concrete and in principle testable prediction - the underlying interchange symmetry $N_1 \leftrightarrow N_2$ implies that one the light left-handed neutrinos is exactly massless, and within the tribimaximal mixing approximation we may have either $m_1\simeq m_2=0$ (normal hierarchy) or $m_3=0$ (inverted hierarchy). Further testing of this scenario may, however, be challenging in the near future, since the seesaw mechanism implies either Majorana masses within the kinematical reach of collider experiments such as the LHC but which couple very weakly to SM particles, or significant Yukawa couplings but too heavy right-handed neutrinos. Nevertheless, the model makes a very concrete prediction that dark matter is made of scalar particles that only coupled directly to two of the right-handed neutrinos and, moreover, couples equally to $N_1$ and $N_2$, albeit with couplings of opposite signs. Hence, even if directly testing the $\nu$IDM model is beyond the reach of current technology, the model should yield distinctive signatures that, hopefully in a not too distant future, may be probed in the laboratory.

We have thus developed one of the simplest and economical extensions of the SM that can simultaneously address the problems of inflation, dark matter, neutrino masses and the cosmological baryon asymmetry, adding only a single scalar field and three Weyl fermions. As mentioned above, it would be interesting, and in some cases even desirable, to embed this construction within a supersymmetric framework, and we hope that this work motivates further exploration of this and possibly other related ideas towards constructing a complete model of the early Universe.

\acknowledgments{ J.G.R. is supported by the FCT Investigator Grant No. IF/01597/2015. This work was partially supported by the H2020-MSCA-RISE-2015 Grant No. StronGrHEP-690904 and by the CIDMA Project No. UID/MAT/04106/2013.
}





\end{document}